\documentstyle[12pt,epsfig]{article}
\let\section=\subsection  \let\subsection=\subsubsection

\def\be{\begin{equation}}
\def\ee{\end{equation}}
\def\bea{\begin{eqnarray}}
\def\eea{\end{eqnarray}}

\begin{document}
\begin{center}
{\large \bf Baryon Spectrum in Large $N_c$ Chiral Soliton\\ and in Quark Models.
}\\[5mm]
Vladimir B. Kopeliovich and Andrei M. Shunderuk   \\[2mm]
{\small \it
Institute for Nuclear Research of RAS, Moscow 117312, Russia}
\end{center}

In memory of {\bf Karen Avetovich Ter-Martirosyan}, the Teacher.\\[2mm]
\begin{abstract}\noindent
\tenrm \baselineskip=11pt
Strangeness contents of baryons are calculated within the rigid rotator model for arbitrary
number of colors $N_c$. The problem of extrapolation to realistic value $N_c=3$ is noted,
based on explicit calculations and comparison of the rigid rotator and rigid oscillator variants
of the model. Some features of exotic baryon spectra ($\{\bar{10}\},\; \{27\}$-and
$\{35\}$-plets of baryons) obtained in the chiral soliton approach can be understood in terms
of simplified quark $(4q\bar{q})$ wave functions.
The effective mass of strange antiquark in different $SU(3)$ multiplets of pentaquarks should depend
on the particular multiplet, to link the predictions of soliton and quark models.
The estimate of the $6_F$ and $\bar{3}_F$ diquarks mass difference can be made from comparison
with chiral soliton model results for masses of exotic baryons from different $SU(3)$-multiplets.
The masses of baryons partners with different values of spin $J$ are also estimated.
\end{abstract}
\bigskip
\leftline{PACS numbers: 12.39.Dc,  12.40.Yx,  14.20.-c,  14.20.Gk}
\vspace{0.1cm}
\baselineskip=13pt
\section{Introduction }
Description of hadrons structure in terms of their quark constituents is generally accepted,
but the alternative description within e.g. topological soliton (Skyrme) model \cite{skyrme,witten}
and its modifications also is useful and has certain advantages in comparison with
traditional approaches. The chiral (topological) soliton approach is based on general principles
and few ingredients incorporated in the effective chiral lagrangian, this is the reason for
apparent simplifications in comparison, for example, with attempts to solve relativistic many-body
problem. To simplify the latter, some additional objects like diquarks and triquarks have been
phenomenologically introduced and discussed especially intensively after recent observations of the 
so called pentaquarks \cite{kl,jw} \footnote{A contradictive present situation with experimental
observation of possible pentaquark states is discussed, e.g. in \cite{hicks,schum}. Consideration of
baryon states in the present paper is relevant independently on particular values of masses, widths and 
other properties of exotic baryon states measured experimentally. A detailed discussion
of theoretical predictions of these states can be found in \cite{pr1,vk05,jenmal,vkuf,did}}.
Concept of diquarks "as an organizing principle for
hadron spectroscopy" is considered in details in \cite{fw}, see also \cite{kl2}.
The concepts of diquarks, triquarks or other correlated quark clusters are certainly of useful
heuristic value, although their properties have not been
deduced rigorously from basic QCD lagrangian.  It should be noted that diquarks present in different physical states,
baryons or mesons, can have different properties like the effective mass and size, even for
same quantum numbers. \footnote{Some analogy with nuclei can be noted: two-, three-, etc.
nucleon clusters play an important role in the structure of heavy nuclei, however, it is not
possible to evaluate their properties from those of deuteron, helium etc., only. See, e.g. \cite{forest}
for discussion of the role of femtometer toroidal structures in nuclei.}

In the present paper we perform explicit calculation of the strangeness contents of exotic and
nonexotic baryon states at arbitrary number of colors $N_c$, and discuss connection of the 
chiral soliton approach (CSA) and simple quark (pentaquark) model for exotic baryon states, 
in the realistic $N_c=3$ case.
Although there was intensive discussion of connections of the rigid rotator model (RRM) and the
bound state model (BSM) in the literature, mainly in the large $N_c$ limit 
\cite{tc,ikor,pr2,dd,pobylitsa,jenman,tcl,hw2},
explicit analytical calculations of observable quantities at arbitrary $N_c$ were lacking still, 
except several cases \cite{vk05,hw2},\cite{tcl}. The rotation-vibration approach (RVA) described 
in \cite{hw2,herbert} and references in these papers, includes both rotational (zero modes) and 
vibrational degrees of freedom of solitons and is generalization of both RRM and BSM, which appear 
therefore as particular variants of RVA when certain degrees of freedom are frozen 
(see also discussion in Section 3)\footnote{The approach of \cite{hw2} was criticized in \cite{cww},
and response to this criticism was given in \cite{hw3}.}. 
As our studies have shown, there is essential difference between results of RR calculation and
BS model (in its commonly accepted version) in the next to leading term contributions of the
$1/N_c$ expansion for the mass splittings inside $SU(3)$ multiplets of baryons. Since the expansion
parameter is large, there is a problem of extrapolation from the large $N_c$ limit to the real $N_c=3$
world. This problem of extrapolation to realistic value of $N_c$ we 
note in the BSM, persists in RVA as well.

Some features of exotic baryon spectra obtained previously within topological soliton model 
\cite{wk,pr1,wuma,ellis,trampetic} can be understood in the framework of 
pentaquark model, independently of its particular variant (see, e.g. \cite{ok}). 
The Gell-Mann --- Okubo relations which are valid in any model
where the $SU(3)$ symmetry breaking is introduced in a definite way, mimic the mass splittings of
simple quark models, where they  are mainly due to the mass difference between
strange and nonstrange quarks.

Comparison of the results of calculation within soliton model - if we believe that CSA provides
correct description, of course -  with naive quark-diquark
model allows us to get information about properties of constituents (quarks, antiquarks, diquarks),
e.g. mass differences of diquarks with different
quantum numbers, in qualitative agreement with other estimates. Quantitatively, however,
the mass difference between "bad" and "good" diquarks obtained in this way contains considerable 
uncertainties. Another result of interest is relatively strong dependence of the mass of 
strange antiquark on the $SU(3)$ baryon multiplet under discussion.
It is shown as well that the partners of baryon resonances with different $J^P$ predicted within
quark models are present also in the CSA, although they have usually higher energy.
Some of these questions have been addressed in talks \cite{vk05}, and here we add more rigour
to this consideration.

In the next section strangeness contents of nonexotic and exotic states are calculated at
an arbitrary number of colors $N_c$, in section 3 these results are compared with that of
the bound state model, its rigid oscillator (RO) variant, for a small enough value of the flavor
symmetry breaking mass. In section 4 comparison with
the simple pentaquark model is performed, the partners of baryon states with different spin are discussed
in section 5, and the final section contains some conclusions.
\section{Strangeness contents of baryons for arbitrary number of colors}
We begin our consideration with scalar strangeness contents ($C_S$ in what follows) of baryons, 
nonexotic and exotic, which defines the mass splittings within $SU(3)$ multiplets of baryons 
in the chiral soliton approach, by following reasons.
First, strangeness content of baryons or baryon resonances is important and a physically transparent
characteristic of these states, not calculated yet analytically for an arbitrary number of colors $N_c$
\footnote{Numerical calculations for the "octet" and "decuplet" of baryons have been performed
recently, however (Herbert Weigel, private communication, see also \cite{hw2}).}.
Second, the behavior of this quantity as function of $N_c$ allows to make some conclusions (mostly 
pessimistic) about the possibility of extrapolation from large $N_c$ to realistic world with $N_c=3$. 
Comparison of different variants of the model at arbitrary $N_c$ and at $N_c\to 3$ also allows to make
conclusions about reliability of the whole CSA.

The spectrum of observed baryon states is obtained within chiral (topological)
soliton models by means of quantization of the motion of starting
classical field configuration (usually it is $SU(2)$ configuration, although it
may be some other configuration as well) in $SU(3)$ collective coordinates space.
In the rigid rotator approximation the mass formula for the quantized states is
\cite{G,Weigel,ksr,wk,ksh}
\be\label{mf}
M(p,q,Y,I,J)=M_{cl} + \left[C_2(SU_3) -J(J+1)-{N_c^2\over 12}\right]{1\over 2\Theta_K} + 
\frac{J(J+1}{2\Theta_\pi} + \Delta M,
\ee
Second order Casimir operator $C_2(SU_3)=(p^2+q^2+pq)/3 +p+q$ for the $(p,q)$-multiplet,
$Y,\,I,\,J$ are the hypercharge, isospin and spin of baryon, $\Theta_\pi$ and $\Theta_K$ are the 
moments of inertia, of the order of $(5-6)\,GeV^{-1}$ and $(2-3)\,GeV^{-1}$.
The mass splittings within multiplets of baryons are defined by the following relation \cite{G}, 
see also \cite{Weigel,ksr,wk,ksh} where details of evaluation and expresiions for the moments 
of inertia can be found:

{\newcommand\rbox[1]{\makebox[106pt][r]{$#1$}}
\begin{center}
\begin{tabular}{|l|l|l|l|l|l|}
\hline
$[p,q]$& $\qquad  \qquad \qquad C_S(N)$ & $C_S(N=3)$ \\
\hline
$[1,(N-1)/2]$& &\\
\hline
$\;Y'=1,\,I=1/2$  & $\rbox{2(N+4)}/[(N+3)(N+7)]$& $\;7/ 30$\\
$\;Y'=0,\,I=0$  & $\rbox3/(N+7)$ &$\;9/ 30$\\
$\;Y'=0,I=1$  & $\rbox{(3N+13)}/[(N+3)(N+7)]$& $11/ 30$\\
$\;Y'=-1,I=1/2$ & $\rbox4/(N+7)$&$ 12/ 30$\\
$*Y'=-1,I=3/2$ & $\rbox{(4N+18)}/[(N+3)(N+7)]$& --- \\
\hline
$[3,(N-3)/2]$& &\\
\hline
$\;Y'=1,\,I=3/2$  & $\rbox{2(N+4)}/[(N+1)(N+9)]$& $\;7/ 24$\\
$\;Y'=0,\,I=1$  & $\rbox{(3N+7)}/[(N+1)(N+9)]$ &$\;8/ 24$\\
$*Y'=0,\,I=2$  & $\rbox{(3N+15)}/[(N+1)(N+9)]$ & ---\\
$\;Y'=-1,\,I=1/2$ & $\rbox{(4N+6)}/[(N+1)(N+9)]$& $\;9/ 24$\\
$*Y'=-1,\,I=3/2$ & $\rbox{4(N+3)}/[(N+1)(N+9)]$& --- \\
$*Y'=-1,\,I=5/2$ & $\rbox{(4N+22)}/[(N+1)(N+9)]$& --- \\
$\;Y'=-2,I=0$ & $\rbox5/(N+9)$&$10/24$ \\
$*Y'=-2,I=1$ & $\rbox{(5N+9)}/[(N+1)(N+9)]$& --- \\
$*Y'=-2,I=2$ & $\rbox{(5N+17)}/[(N+1)(N+9)]$& --- \\
$*Y'=-2,I=3$ & $\rbox{(5N+29)}/[(N+1)(N+9)]$& --- \\
\hline
$[0,(N+3)/2]$& &\\
\hline
$\;Y'=2,\,I=0$ & $\rbox3/(N+9)$& $\;6/ 24$\\
$\;Y'=1,\,I=1/2$ & $\rbox{(4N+9)}/[(N+3)(N+9)]$ &$\;7/ 24$\\
$\;Y'=0,\,I=1$ & $\rbox{(5N+9)}/[(N+3)(N+9)]$& $\;8/ 24$\\
$\;Y'=-1,\,I=3/2$& $\rbox{(6N+9)}/[(N+3)(N+9)]$&$ \;9/ 24$\\
$*Y'=-2,\,I= 2$& $\rbox{(7N+9)}/[(N+3)(N+9)]$& --- \\
\hline
\end{tabular}
\end{center} }
{\bf Table 1.} {\tenrm Strangeness contents of the "octet", "decuplet" and "antidecuplet" of
baryons at arbitrary $N=N_c$, for unmixed states. $Y'=S+1$, states which appear only if $N>3$ are 
marked by $*$.}\\
\\
\be
\label{DM}
\Delta M= \left[ \Gamma \left({F_K^2\over F_\pi^2}\mu_K^2-\mu_\pi^2\right)+
\bigl(F_K^2-F_\pi^2\bigr)\tilde{\Gamma}\right] C_S,
\ee
if the configuration mixing is not included.

\be\label{sigma}
\Gamma = {F_\pi^2\over 2} \int (1-c_f) d^3r
\ee
is so called $\sigma$ term, one of characteristics of the classical configuration, and
\be
\tilde{\Gamma}={1\over 4}\int c_f\biggl(f'^2+{2s_f^2\over r^2}\biggr) d^3r,
\ee
 $f$ is the profile function of the skyrmion,  the values of physical masses
 $\mu_K,\;\mu_\pi$ and decay constants $F_K,\;F_\pi$ are taken from experiment.
 $C_S$ is so called strangeness content
 of the quantized baryon state.
Within the RR model \cite{G} rotation of incident $SU(2)$ configuration is
described with the help of matrix of collective coordinates $A(t)\in SU(3)$,
which is usually parameterized as
$A= A_1(SU_2) exp(i\nu\lambda_4) A_2(SU_2) exp(i\rho\lambda_8/\sqrt 3)$,
where $SU(2)$ rotation matrices depend each of 3 variables, $\lambda_4,\lambda_8$
are Gell-Mann matrices, see e.g. \cite{deSwart}.

The wave functions of baryons in $SU(3)$ space, $\Psi (p,q;Y,I,I_3)$ are just
$SU(3)$ Wigner functions
depending on 8 parameters incorporated in matrix $A$: the integers $(p,q)$
define the $SU(3)$ multiplet under consideration, $Y,I,I_3$ are hypercharge,
isospin and its 3-d projection of particular baryon state. For arbitrary (odd) number of
colors the hypercharge is connected with strangeness by relation $Y=S+N B/3$, see e.g. 
\cite{witten,tc,vk05} (we omit the index $c$ in $N_c$ in most of formulas and in Tables 1,2). 
It is more convenient therefore to use for the $B=1$ case the quantity $Y'=S+1$, 
as we do in Tables and in Fig. 1. Within this
parametrization the only flavor changing parameter is $\nu$, which defines the
deviation in "strange direction", and strangeness content
\be
C_S={1\over 2}<\Psi_B(\nu)|sin^2\nu|\Psi_B(\nu)>,
\ee
see Appendix, where explicit examples of $\nu$-dependent wave functions of some baryon states
are given.
The main contribution to the baryon mass operator, depending on flavor symmetry
breaking (FSB) mass $m_K$ equals to \cite{G}
\be \label{dmrr}
\Delta M = m_K^2 \Gamma {<1- D_{88}(\nu)>\over 3} = {1\over 2}m_K^2\  \Gamma\  <sin^2\nu>,
\ee
since $D_{88}=Tr (A^\dagger \lambda_8 A \lambda_8)/2 = 1-3 (sin^2\nu)/2$,
$m_K^2 = F_K^2 \mu_K^2/F_\pi^2 - \mu_\pi^2$. 

Second term in (\ref{DM}), proportional to $\tilde{\Gamma}$ gives relatively small contribution
in comparison with the first term; it is however not negligible for realistic values of
masses and parameters.
When $m_K \to 0$, then the $\nu$ rotation becomes zero mode.
More details can be found e.g. in \cite{wk,vk05,ksh}.

The quantity $<D_{88}>$ can be calculated using Clebsh-Gordan coefficients for arbitrary number
of colors $N$ which have been presented previously for few cases in \cite{kk,tcl}  (however, the
strangeness contents have not been calculated).
Another method of calculations which we prefer here is
to use the baryons wave functions in the $SU(3)$-configuration space, as it was described, e.g.
in \cite{Weigel}.  For large $N$ generalization of the octet, the decuplet of
baryons and for the $\Theta^+$ baryon this method has been used
recently in \cite{vk05} to calculate strangeness contents of these baryons.
For exotic baryon multiplets, "antidecuplet", $"\{27\}"$- and
$"\{35\}"$-plets (shown in Fig. 1) we present here strangeness contents and wave functions for the 
first time (see Appendix).
In Tables 1,2 strangeness contents are given for arbitrary number of colors, and also
numerically for $N=3$.

It can be seen easily from Tables 1 and 2 that for the fixed value of strangeness, $C_S$ decreases
as $1/N$ with increasing $N$ - in agreement with the fact that fixed number of quarks are strange,
whereas total number of constituent quarks is $N$, or $N+2$ for "pentaquarks".
The difference of strangeness contents of states from different
$SU(3)$ multiplets, but with the same value of strangeness,  decreases as $1/N^2$ or faster.
E.g., the difference of $C_S$ for the "nucleon" with $I=1/2$ and "delta" ($I=3/2$) decreases like
$1/N^3$ \cite{vk05}.
\begin{figure}
\label{multiplet}
\begin{center}
\epsfig{figure=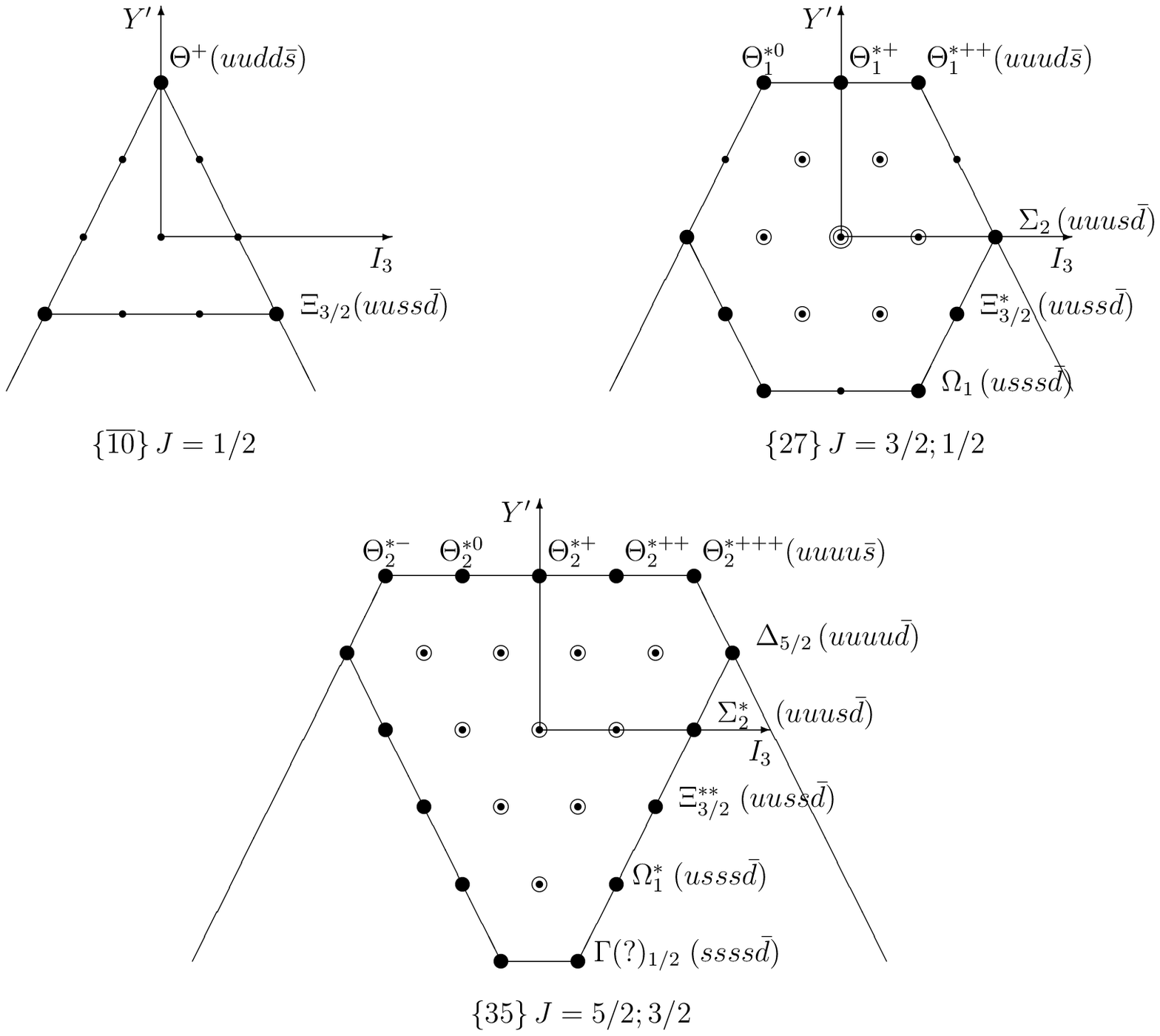,width=16cm,angle=0}
\vspace{1.5cm}
\protect\caption{\tenrm
The $I_3-Y'$ diagrams $(Y'=S+1)$ for multiplets of pentaquark baryons, antidecuplet,
$\{27\}$- and $\{35\}$-plets. For $N>3$ these diagrams should be extended within long
lines, as shown in the picture. Quark contents are given for manifestly exotic states, when
$N=3$.}
\end{center}
\end{figure}

\newcommand\rbox[1]{\makebox[106pt][r]{$#1$}}
\begin{center}
\begin{tabular}{|l|l|l|l|l|l|}
\hline
$[2,(N+1)/2]$&$\qquad  \qquad \qquad C_S(N)$ & $C_S(N=3)$\\
\hline
$\;Y'=2,\,I=1$ & $\rbox{(3N+23)}/[(N+5)(N+11)]$& $32/ 112$\\
\hline
$\;Y'=1,\,I=3/2$ & $\rbox{(4N^2+65N/2-3/2)}/[(N+1)(N+5)(N+11)]$& $33/ 112$\\
$\;Y'=1,\,I=1/2$ & $\rbox{(4N+24)}/[(N+5)(N+11)]$& $36/ 112$\\
\hline
$\;Y'=0,\,I=2$ & $\rbox{(5N^2+39N-26)}/[(N+1)(N+5)(N+11)]$ & $34/ 112$\\
$\;Y'=0,\,I=1$ & $\rbox{(5N^2+33N+8)}/[(N+1)(N+5)(N+11)]$ & $38/ 112$\\

$\;Y'=0,\,I=0$ & $\rbox{5}/(N+11)$ & $5/ 14$\\
\hline
$*Y'=-1,\,I=5/2$ & $\rbox{(6N^2+{91\over 2}N-{101\over 2})}/[(N+1)(N+5)(N+11)]$ & --- \\
$\;Y'=-1,\,I=3/2$ & $\rbox{(6N^2+38N-8)}/[(N+1)(N+5)(N+11)]$ & $40 / 112$\\
$\;Y'=-1,\,I=1/2$ & $\rbox{(6N+7/2)}/[(N+1)(N+11)]$ & $43/112$\\
\hline
$*Y'=-2,\,I=3$& $\rbox{(7N^2+52N-75)}/[(N+1)(N+5)(N+11)]$& ---\\
$*Y'=-2,\,I=2$& $\rbox{(7N^2+43N-24)}/[(N+1)(N+5)(N+11)]$& ---\\
$\;Y'=-2,\,I=1$& $\rbox{(7N+2)}/[(N+1)(N+11)]$& $46/ 112$\\
\hline
$[4,(N-1)/2]$& &\\
\hline
$\;Y'=2,\,I=2$ & $\rbox{(3N+25)}/[(N+3)(N+13)]$& $34/96$\\
\hline
$\;Y'=1,\,I=5/2$ & $\rbox{(4N^2+85N/3-79)}/[(N-1)(N+3)(N+13)]$ & $21/96$\\
$\;Y'=1,\,I=3/2$ & $\rbox{(4N+24)}/[(N+3)(N+13)]$ & $36/96$\\
\hline
$*Y'=0,\,I=3$ & $\rbox{(5N^2+{104\over 3}N-133)}/[(N-1)(N+3)(N+13)]$ & ---\\
$\;Y'=0,\,I=2$   & $\rbox{(5N^2+{74\over 3}N-67)}/[(N-1)(N+3)(N+13)]$ & $26/96$\\
$\;Y'=0,\,I=1$& $\rbox{(5N+23)}/[(N+3)(N+13)]$ & $38/96$\\
\hline
$*Y'=-1,\,I=7/2$& $\rbox{(6N^2+41N-187)}/[(N-1)(N+3)(N+13)]$ & ---\\
$*Y'=-1,\,I=5/2$& $\rbox{(6N^2+{88\over 3}N-110)}/[(N-1)(N+3)(N+13)]$ & ---\\
$\;Y'=-1,\,I=3/2$& $\rbox{(6N^2+21N-55)}/[(N-1)(N+3)(N+13)]$ & $31/96$\\
$\;Y'=-1,\,I=1/2$& $\rbox{(6N+22)}/[(N+3)(N+13)]$ &$40/96$ \\
\hline
$*Y'=-2,\,I=4$& $\rbox{(7N^2+{142\over 3}N-241)}/[(N-1)(N+3)(N+13)]$ & ---\\
$*Y'=-2,\,I=3$& $\rbox{(7N^2+34N-153)}/[(N-1)(N+3)(N+13)]$ & ---\\
$*Y'=-2,\,I=2$& $\rbox{(7N^2+24N-87)}/[(N-1)(N+3)(N+13)]$ & ---\\
$\;Y'=-2,\,I=1$& $\rbox{(7N^2+52N/3-43)}/[(N-1)(N+3)(N+13)]$ & $36/96$\\
$\;Y'=-2,\,I=0$& $\rbox{7}/(N+13)$ & $42/96$\\
\hline
$*Y'=-3,\,I=9/2$& $\rbox{(8N^2+{161\over 3}N-295)}/[(N-1)(N+3)(N+13)]$&---\\
$*Y'=-3,\,I=7/2$& $\rbox{(8N^2+{116\over 3}N-196)}/[(N-1)(N+3)(N+13)]$& ---\\
$*Y'=-3,\,I=5/2$& $\rbox{(8N^2+27N-119)}/[(N-1)(N+3)(N+13)]$& ---\\
$*Y'=-3,\,I=3/2$& $\rbox{(8N^2+56N/3-64)}/[(N-1)(N+3)(N+13)]$          & ---\\
$\;Y'=-3,\,I=1/2$& $\rbox{(8N-31/3)}/[(N-1)(N+13)]$             & $41/96$\\
\hline
\end{tabular}
\end{center}
 {\bf Table 2.} {\tenrm Strangeness contents for unmixed states of the $"\{27\}"$-plet
 (spin $J=3/2$) and
 $"\{35\}"$-plet ($J=5/2$) of baryons, for arbitrary $N$ and  numerically for $N=3$.
 States which exist only for $N>3$ are marked with $*$.}\\

Any chain of states within definite $SU(3)$ multiplet, satisfying relation $I=\pm Y/2 +\,C$,
i.e. which belong to such straight lines in $(I-Y')$--- plane, has equidistant behavior due
to Gell-Mann --- Okubo relations \footnote{The validity of Gell-Mann --- Okubo relations for the
"octet" and "decuplet" of baryons at an arbitrary number of colors has been noted long ago in
the paper \cite{djm} where the $1/N$ expansion and induced representation methods were
developed for describing baryon properties.}. According to these, the mass splitting and 
strangeness contents within the $SU(3)$ multiplets can be presented in the form
\be\label{gmo}
C_S(p,q,Y',I) = a(p,q)Y'+b(p,q)[Y'^2/4-I(I+1)]+c(p,q),
\ee
where $a(p,q),\;b(p,q)$, being constants within any $SU(3)$ multiplet, are different for different
multiplets $(p,q)$. Linear behaviour of masses of any chain
of states with $I=\pm Y'/2 +\,C$ follows then immediately. Since $Y'=S+1$, (\ref{gmo}) can be easily
rewritten in terms of strangeness $S$ and isospin $I$.

From Table 1 we easily obtain
\bea\label{abc}
a("\{8\}") &=& -\,{N+2\over (N+3)(N+7)},\quad b("\{8\}")=-\,{2\over (N+3)(N+7)},\nonumber\\
 c("\{8\}")&=&{3\over (N+7)},
\eea
and for "decuplet":
\bea\label{abc10}
a("\{10\}")&=&-\,{N+2\over (N+1)(N+9)},\quad b("\{10\}")=-\,{2\over (N+1)(N+9)},\nonumber\\
c("\{10\}")&=&{3\over (N+9)}.
\eea
For "antidecuplet"  $I=1-Y'/2$, relation (\ref{gmo}) takes the form
$$ C_S= (a+3b/2)Y' -2b+c, $$ and we obtain from Table 1 two relations:
\bea\label{abca10}
a(\{"\overline{10}\}")+{3\over 2}b("\{\overline{10}\}")&=& -{N\over (N+3)(N+9)}, \nonumber\\
-2b(\{"\overline{10}\}")+c("\{\overline{10}\}") &=& {5N+9 \over (N+3)(N+9)}.
\eea
For $"\{27\}"$-plet we have from Table 2:
\bea
a("\{27\}")&=&{-(N^2+11N/4-13/4)\over (N+1)(N+5)(N+11)},\; b("\{27\}")={-(3N-17)\over 
2(N+1)(N+5)(N+11)},\nonumber\\
c("\{27\}")&=&{5\over (N+11)}.
\eea
and for $"\{35\}"$-plet:
\bea
a("\{35\}")&=&{-(N^2+N/2-31/2)\over (N-1)(N+3)(N+13)},\; b("\{35\}")={-(5N/3-11)\over 
(N-1)(N+3)(N+13)},\nonumber\\
c("\{35\}")&=&{5N^2+44N/3-1\over (N-1)(N+3)(N+13)}.
\eea
In all cases at large $N$,  $a(p,q)\sim c(p,q)\sim 1/N$, and $b(p,q)\sim 1/N^2$.
A feature of interest is that the step in $C_S$ per unit strangeness for "decuplet",
$\delta_{10}=(N-1)/[(N+1)(N+9)]$, is greater than that for "antidecuplet",
$\delta_{\overline{10}}=N/[(N+3)(N+9)]$, although they coincide for $N=3$, and we do not consider the
case of $N=1$ 
\footnote{It should be mentioned that it is a convention to identify the multiplet
$[p,q]=[3,(N-3)/2]$ with the "decuplet". In this case the difference $Y^{max}-Y^{min}=p+q=(N+3)/2$ 
coincides
with that of "antidecuplet" $[0,(N+3)/2]$. It is usually assumed for generalization of any $SU(3)$
multiplet that spin and isospin of baryon state is fixed when number of colors $N_c$ increases. Another 
logical possibility for generalization of decuplet, based on symmetry principle,
is the multiplet $[N,0]$, see e.g. discussion in \cite{vk05}.}.

It can be seen also from Tables 1,2 that the parameter for expansion $C_S=(\alpha/N)[1+\beta/N+...]$
is $\sim 7/N, \;9/N,\; 11/N,\; 13/N,...$, for the "octet", "decuplet", $"\{27\}"$ and $"\{35\}"$-plets,
so, it increases with increasing values of $(p,q)$ defining the multiplet \cite{vk05}. E.g., for
multiplets $[p,q]=[0,(N+3m)/2]$ the expansion parameter is $(3m+6)/N$. The authors of \cite{hw2} 
came to
similar conclusions considering the decay matrix element for $\Theta^+$-baryon: "Any approach ...
that employs $1/N$ expansion methods for exotic baryon matrix elements seems questionable"
(subsection VI.B of \cite{hw2}). As we show here, for nonexotic baryons such expansion method
is questionable also, for the bound state model as well as for the RVA.
\section{Comparison of rigid rotator and oscillator models at large $N$}
When flavor symmetry breaking mass $m_K$ is small enough, it is possible to compare directly
results of the rigid rotator and oscillator models at arbitrary $N$. In the RR model any baryon
state is ascribed, at first, to definite $SU(3)$-multiplet $(p,q)$ with some value of spin $J$
which depends on the multiplet, and as a next step the mass splitting within each multiplet
can be calculated in the first order in FSB mass $m_K$, precisely for arbitrary number of colors $N$ 
(previous section). In the bound state model \cite{ck,kk,westk} expansion in $1/N$ is made from
the beginning, the states are labeled by their strangeness (flavor in general case), spin and 
isospin. The $J,\,I$ - dependent energy is calculated as the hyperfine splitting correction of the
order $\sim 1/N$, and each state can be ascribed to definite $SU(3)$-multiplet, according
to its quantum numbers $S,\,I$ and $J$. When $m_K \to 0$,
there is no need to consider the full bound state model, because it reduces in this limit
to simplified rigid oscillator version \cite{westk,kwest}.
\subsection{Nonexotic baryon states}
In this subsection we follow mainly to discussion in \cite{kleko}.
For the rigid rotator model we shall use the above expressions (\ref{DM}) - (\ref{dmrr}), i.e.
\be
\Delta M = m_K^2\Gamma C_S,
\ee
which corresponds to first order in flavor symmetry breaking mass squared $m_K^2$. This approximation
becomes more precise as $m_K^2 \to 0$. In this limit the RR model and soft, or slow rotator
model provide same results \footnote{The opposite to the rigid rotator is the assumption that 
during the rotation it is sufficient
time for changing the skyrmion profiles under influence of FSB terms in
the lagrangian (so called soft, or slow rotator approximation, see
\cite{Schwesinger} where static properties of baryons have been calculated
within this approximation). Evidently, both rigid and soft rotator approximations
converge when $m_K \to 0$, and estimates show also that for $B=1$ the
RR model is more justified in the realistic case, whereas for large baryon
numbers the soft rotator model can be better \cite{vk05}.}.

From Table 1 we obtain for the components of the "octet", providing expansion in parameter $1/N$: 
\be\label{dmn}
\delta M_N =  {2 (N+4)\over (N+3)(N+7)} m_K^2 \Gamma
= \left ({2\over N} -{12\over N^2} + O(N^{-3})\right )
m_K^2 \Gamma
\ ,
\ee
\be
\delta M_\Lambda =  {3 \over (N+7)} m_K^2 \Gamma
= \left ({3\over N} -{21\over N^2} + O(N^{-3})\right )
m_K^2 \Gamma
\ ,
\ee
\be
\delta M_\Sigma =  {3N+13\over (N+3)(N+7)} m_K^2 \Gamma
= \left ({3\over N} -{17\over N^2} + O(N^{-3})\right )
m_K^2 \Gamma
\ ,
\ee
\be\label{dmxi}
\delta M_\Xi =  {4 \over (N+7)} m_K^2 \Gamma
= \left ({4\over N} -{28\over N^2} + O(N^{-3})\right )
m_K^2 \Gamma
\ .
\ee
For arbitrary nonexotic $SU(3)$ multiplets it is a matter of simple algebra to
show, using the $\nu$-dependent wave functions of baryons, that for
not large values of $S$ the strangeness content of baryon equals to
\be \label{anymult}
 C_S \simeq {2+|S| \over N},
\ee
so, minimal strangeness content exists and decreases like $1/N$.

Let us compare this with the results of the RO approach.
The bound state soliton model is in fact particular case of the more general rotation-vibration
approach (RVA) described in details in \cite{hw2}, see also references in this paper.
In the rigid oscillator model parametrization of the matrix $A(t)$ is
used, somewhat different from that described above: $A(t)=A_{SU(2)}(t)S(t)$, matrix
$S(t)=exp(i {\cal D})$ describes strangeness changing movement of soliton in $SU(3)$ space
\cite{ck,kk}:
\be
{\cal D} = \sum_{a=4}^7 d_a\lambda_a,
\ee
so, deviation into "strange" direction is defined by two-component spinor
$D=(d_4-id_5,d_6-id_7)^T/\sqrt 2$. Comparison with the RR parametrization
above allows to conclude that $D^\dagger D \simeq \nu^2/2$. The hamiltonian
of RO model is of the oscillator type and can be quantized appropriately
\cite{kk,westk}.
The average deviation $|D|$ into strange direction for arbitrary negative $S$ can be 
estimated easily as
\be\label{dev}
|D|_S \sim \frac{2|S|+1}{\big [16 m_K^2 \Gamma \Theta_K + N^2 \big ]^{1/4}},
\ee
for $S<0$. At fixed $|S|$ it decreases with increasing $N$ and FSB mass $m_K$.
However, (\ref{dev}) does not hold for positive $S$.
The quantity $\Theta_K$, similar to $\Gamma$, is defined by incident $SU(2)$
chiral field configuration \cite{kk,westk}, and can be called
the moment of inertia of skyrmion relative to the motion into "strange"
direction. It is assumed again that during the motion in the oscillator
potential the classical configuration does not change its form, that is
the reason why the model is called the rigid oscillator one.

The order $N^0$ contributions to the non-exotic baryon masses are
\be \label{orderone}
\Delta M_0(RO)= \omega_- + \omega_+ + \omega_- |S|
\ee
where
\be
\omega_\pm = {N\over 8\Theta_K} (\mu\pm 1)
\ ,
\ee
\be\label{mu}
\mu= \sqrt{1+ (m_K/M_0)^2}\ , \qquad M_0 = {N\over 4\sqrt{\Gamma\Theta_K} }
\ .
\ee
In lowest order in $m_K$ we obtain easily:
\be
\omega_-\simeq m_K^2{\Gamma\over N}, \qquad
\omega_+\simeq {N\over 4\Theta_K} +m_K^2{\Gamma\over N}.
\ee
The first two terms in (\ref{orderone}) come from the zero-point energy.
To order $m_K^2$ this gives
\be \label{dm2}\Delta M_0(RO) \simeq {N\over 4\Theta_K} + {m_K^2\Gamma\over N} (2 + |S|)
\ ,
\ee
The term $N/(4\Theta_K)$ is well known to appear in the RR
approach \cite{pr1,wk,vk05}, and we also see that the term linear in $m_K^2$ agrees with the RR
approach, in the order $N^0\sim 1$.

The $O(1/N)$ contributions were studied in \cite{kk,westk},
and the result was expressed in terms of the hyperfine splitting (HFS) constants
\be
 c = 1- {\Theta_\pi\over 2\mu\Theta_K} (\mu-1)=
1- {4\Theta_\pi \Gamma m_K^2\over N^2} + O(m_K^4)\ ,
\ee
\be
\bar c = 1- {\Theta_\pi\over \mu^2\Theta_K} (\mu-1)=
1- {8\Theta_\pi \Gamma m_K^2\over N^2} + O(m_K^4)\ .
\ee
The $O(1/N)$ term as stated in \cite{westk} and obtained also in \cite{ksr,ksh}, is
\be\label{EHFS}
\Delta E_{HFS}={J(J+1)\over 2\Theta_\pi}+ {1\over 2\Theta_\pi}\left \{(c-1)
\left [J(J+1) -I(I+1)\right] +(\bar c-c)I_S\left (I_S+1)\right ) \right \}
\ .
\ee
with $I_S=|S|/2$ - isospin carried by kaon field 
\footnote{In \cite{kk,westk} the last term in the bracket of (\ref{EHFS}) was given
as $(\bar c -c)Y^2/4$, correct formula was given first in \cite{kwest} and for general case in
\cite{ksr}. Details of the evaluation can be found also in \cite{ksh}.}.
At $m_K=0$ (flavor symmetric case) $c=\bar c=1$, and the hyperfine splitting correction reduces
to the well known quantum rotational correction $J(J+1)/ 2\Theta_\pi$.
The relations take place in the linear in $m_K^2$ approximation:
\be
\bar c \simeq 2c-1,
\ee
which ensures validity of the Gell-Mann --- Okubo relations, and
\be \label{cc}
\bar c \simeq c^2,
\ee
which is used sometime in literature.
However, relation (\ref{cc}) does not hold for antiflavor (positive strangeness), see next
subsection. In the expression (\ref{EHFS}), the term linear in $m_K^2$ is found to be
\be\label{dmro}
\delta M_{1/N}(RO)= 2{\Gamma m_K^2\over N^2} \left [ I(I+1)- J(J+1)- I_S(I_S+1) \right ]
\ ,
\ee
and for $J=1/2$ we can compare this with the RR results for the "octet", (\ref{dmn}-\ref{dmxi}).
 Collecting the terms $\sim m_K^2\Gamma$ from (\ref{dm2}) and (\ref{dmro}) we obtain
\bea \label{dmroo}
\delta M_N(RO)  \simeq {2\over N} m_K^2\Gamma; \qquad \qquad \qquad
\delta M_\Lambda (RO) &\simeq& \left({3\over N}-{3\over N^2}\right) m_K^2\Gamma; \nonumber \\
\delta M_\Sigma (RO)\simeq \left({3\over N}+{1\over N^2}\right) m_K^2\Gamma; \qquad 
\delta M_\Xi (RO)&\simeq& \left({4\over N}-{4\over N^2}\right) m_K^2\Gamma.
\eea
Obviously, there is no agreement between (\ref{dmroo}) and 
(\ref{dmn}-\ref{dmxi}) for all 4 components of the "octet".

Now, let us consider the ``decuplet'' of baryons, i.e. the
$(3, (N-3)/2)$ multiplet of $SU(3)$, $J=3/2$. The terms linear in
$m_K^2$ as it follows from Table 1, are
\be\label{dmd}
\delta M_\Delta =  {2 (N+4)\over (N+1)(N+9)} m_K^2 \Gamma
= \left ({2\over N} -{12\over N^2} + O(N^{-3})\right )
m_K^2 \Gamma
\ ,
\ee
\be
\delta M_{\Sigma^*} =  {3N+7\over (N+1)(N+9)} m_K^2 \Gamma
= \left ({3\over N} -{23\over N^2} + O(N^{-3})\right )
m_K^2 \Gamma
\ ,
\ee
\be
\delta M_{\Xi^*} =  {2(2N+3) \over (N+1)(N+9)} m_K^2 \Gamma
= \left ({4\over N} -{34\over N^2} + O(N^{-3})\right )
m_K^2 \Gamma
\ .
\ee
\be\label{dmo}
\delta M_\Omega =  {5 \over (N+9)} m_K^2 \Gamma
= \left ({5\over N} -{45\over N^2} + O(N^{-3})\right )
m_K^2 \Gamma
\ .
\ee
They satisfy the usual equal splitting rule for decuplet, with the splitting
\be
{N-1\over (N+1)(N+9)} m_K^2 \Gamma
= \left ({1\over N} -{11\over N^2} + O(N^{-3})\right )
m_K^2 \Gamma
\ .
\ee
Within the RO variant we should use (\ref{dmro}) with $J=3/2$ and $I=J-I_S$, and obtain in
this way for the components of "decuplet" quite different results.

Possible way to remove disagreement between the RR model and RO variant of the
bound state model is the following \cite{kleko}.
The RO calculation involves normal-ordering ambiguities in quartic terms,
which can correct the overall shift of masses and the term linear
in strangeness that already appeared in the leading order in $1/N$.
Let us assume that the normal ordering corrections change the $O(1/N)$ mass 
formula  by an extra additive term \cite{kleko}
\be\label{dmno}
\Delta M (norm.ord.)= -6 {\Gamma m_K^2\over N^2} (2+ |S|)
\ ,
\ee
which is proportional to the order 1 contribution, but
is down by a power of $N$.

Then, the $O(1/N)$ term in the mass formula becomes \cite{kleko}
\be \label{masscorr}
\delta M(RO, norm.ord.)={\Gamma m_K^2\over N^2} \left [-12+ 2I(I+1)- 2J(J+1)- {S^2\over 2} - 7 |S|
\right ]
\ ,
\ee
and the $O(m_K^2\Gamma/N^2)$ terms of the RO approach agree with the RR
calculations for all the ``octet'' and "decuplet" masses. 

These results show that there should be a specific normal-ordering
prescription that brings the two approaches in complete agreement \cite{kleko}.
As it is well known \cite{ikor,hw2}, in the large $N$ limit both RR and RO approaches coincide.
But the next to leading order corrections in the $1/N$-expansion are large, including the normal ordering
correction, so the problem of extrapolation to real world with $N=3$ cannot be solved by means
of $1/N$ expansion. It should be noted also that besides the $1/N$ corrections we discussed here
there can be also corrections of other types, e.g. corrections of "dynamical"  nature to static
characteristics of skyrmions. By this reason, even if the proper way to remove the difference between 
RR and RO models is found, it may not mean that the whole problem of extrapolation to real $N=3$
world is resolved.
\subsection{Positive strangeness states.}
To calculate the HFS correction in this case, the substitution $\mu \to -\mu$ should be made in
above expressions for the HFS constants $c$ and $\bar c$, and we have in this case:
\be
 c_{\bar S} = 1- {\Theta_\pi\over 2\mu\Theta_K} (\mu+1)=1-{\Theta_\pi\over \Theta_K}+
 {8\Theta_K\Gamma m_K^2\over N^2} + O(m_K^4)
\ee
and
\be
\bar c_{\bar S} = 1+{\Theta_\pi\over \mu^2\Theta_K} (\mu+1)=1+{2\Theta_\pi\over \Theta_K}-
{24\Theta_K\Gamma m_K^2\over N^2} + O(m_K^4)
\ee
In the difference from negative strangeness case, for positive strangeness (antiflavor in
general case) the constants $c\neq 1$ at $m_K=0$, and approximate equality $\bar c \simeq c^2$
is strongly violated now.
For the energy of states with antiflavor we have from (\ref{EHFS})
\bea\label{hfsa}
\Delta E_{HFS+FSB}&=&{J(J+1)\over 2\Theta_\pi}+ {1\over 2\Theta_K}\left [-J(J+1) +I(I+1)
 +3I_S(I_S+1) \right ] + \nonumber \\
          &+& {m_K^2\Gamma\over N^2}\left\{3N-2\left [-J(J+1) +I(I+1)+7I_S(I_S+1) \right ]\right\}
\eea
The case of exotic $S=+1$ states is especially interesting. In this case $I_S=1/2,\;J=I+1/2$, and 
within the RO model we obtain, using the expressions for $c_{\bar S}$ and $\bar c_{\bar S}$:

\be \label{T0}
M_{\Theta_0,J=1/2}= {2N+3\over 4\Theta_K} +{3\over 8\Theta_\pi} +  m_K^2 \Gamma
\biggl({3\over N} - {9\over N^2}\biggr) +M_{cl},\ee

\be \label{T1}
M_{\Theta_1^*,J=3/2}= {2N+1\over 4\Theta_K} +{15\over 8\Theta_\pi} +  m_K^2 \Gamma
\biggl({3\over N} - {7\over N^2}\biggr)+M_{cl},\ee

\be \label{T2}
M_{\Theta_2^*,J=5/2}= {2N-1\over 4\Theta_K} +{35\over 8\Theta_\pi} +  m_K^2 \Gamma
\biggl({3\over N} - {5\over N^2}\biggr)+M_{cl}. \ee
The terms of zero's order in $m_K$ coincide exactly with those given above by RR mass formula 
(\ref{mf}) applied to exotic multiplets $\{\overline{10}\},\,J=1/2$, $\{27\},\,J=3/2$ and 
$\{35\},\,J=5/2$.
As it was expected, there is additional contribution $N/(4\Theta_K)$ to the energy of exotic
states compared with nonexotic states, in agreement with the RR model result 
\footnote{It was shown explicitly in \cite{ksh} 
(formula (52) in Appendix) that within
the RR model the energy difference between exotic and nonexotic baryon states (\ref{dm2}) 
due to difference of corresponding Casimir operators equals to $\Delta E=(NB+3)/(4\Theta_K)$ 
for arbitrary odd $B$. Note that
if the expression for $\Delta E_{HFS}$ (\ref{EHFS}) is used with the term $(c^2-c)I_S(I_S+1)$ 
instead of $(\bar c - c)I_S(I_S+1)$, as sometime in the literature, then results of RR model 
cannot be reproduced correctly within BSM. }.
Let us compare this with the mass splitting correction $\sim m_K^2$, obtained within the RR model, 
see Tables 1,2:
\be
\delta M_{\Theta_0} \simeq m_K^2\Gamma \biggl({3\over N} - {27\over N^2}\biggr), \ee

\be
\delta M_{\Theta_1^*} \simeq m_K^2\Gamma \biggl({3\over N} - {25\over N^2}\biggr), \ee

\be
\delta M_{\Theta_2^*} \simeq m_K^2\Gamma \biggl({3\over N} - {23\over N^2}\biggr). \ee

There is considerable difference between RR and RO models in FSB terms, proportional
to $m_K^2$. This difference can be eliminated if the contribution given by (\ref{masscorr})
\be
\label{DMIK}
\Delta M(norm.ord.,S=1) = -18 m_K^2 {\Gamma\over N^2}
\ee
is added to the RO result, similar to the case of the "octet" and "decuplet" of baryons considered
in \cite{kleko} and in previous subsection. Evidently, the difference between RR and RO models
should be kept in mind, when comparison of predictions of both variants is made. However, in 
the literature discussing relevance of the pentaquarks predictions within CSA this difference was not 
taken into account.

Other states with values of strangeness different from $S=1$ which could be ascribed to exotic
multiplets can be considered similarly, but it is technically more complicated problem.
\subsection{Comparison of the total mass splittings}
Also, it is more difficult to calculate the total mass splittings, especially for 
exotic $SU(3)$ multiplets in RO model.
An important restriction for the whole mass splitting of any $SU(3)$ multiplet follows from
expression (\ref{DM}), since $s_\nu^2 \leq 1$:
\be \label{DM2}
\Delta^{tot} M \leq {1\over 2}\left({F_K^2\over F_\pi^2}\mu_K^2 - \mu_\pi^2\right) \Gamma.
\ee
This restriction is useful for the comparison of different quantization schemes.

Within the RR model it is convenient to use the Gell-Mann --- Okubo formulas (\ref{gmo}),
substituting in this formula $Y^{max}=(p+2q)/3,\, I(Y^{max})=p/2$, and 
$Y^{min}=-(q+2p)/3,\, I(Y^{min})=q/2$ (recall that $Y=N/3 + S$ for arbitrary number of colors).

For "decuplet" $[p,q]=[3,(N-3)/2]$ from (\ref{abc10}) we obtain
\be\label{t10RR}
\Delta^{tot}_{RR}(10) = m_K^2\Gamma {N^2+4N-15\over 2(N+1)(N+9)} \simeq {m_K^2\Gamma \over 2}
\left(1-{6\over N} +{36\over N^2}\right)
\ee

Within the RO model, for any multiplet $(p,q)$ the total mass splitting in the leading in $1/N$
approximation is given by
\be
\Delta^{tot} M(p,q) = \Delta Y \omega_- \simeq m_K^2{\Gamma\over N} (p+q).
\ee
It turned out that in this approximation for $N=3$ the total mass splitting within decuplet is 
8 times greater than within rigid rotator approximation (\ref{t10RR}), for the octet the 
difference is 4 times, as noted already in \cite{vk05}.

The hyperfine splitting correction can be calculated with the help of formula (\ref{masscorr}),
where for "decuplet" we should take $J=3/2$, $I=3/2$ for $S=0$ and $I=(N-3)/4$ for $S=-(N+3)/2$.
Then we obtain
\be\label{t10RO}
\Delta^{tot}_{RO}(\{10\}) = {m_K^2\Gamma \over 2}
\left(1-{6\over N} +...\right)
\ee
in agreement with first two terms in the $1/N$ expansion of above formula (\ref{t10RR}).
Note, that it would be no agreement without addition of special normal ordering contribution
(\ref{dmno}) \cite{kleko}. However, there is no agreement in the next order terms in the 
$1/N$ expansion. Of course, one should not expect such agreement because the RO model we are
using here does not take into account such contributions.
Similar agreement between RR and RO results takes place for the total mass splitting of the 
"octet" $[p,q]=[1,(N-1)/2]$.

Let us consider as the next example the "antidecuplet" $[p,q]=[0,(N+3)/2]$ multiplet which is 
generalization of $(0,3)$ antidecuplet for arbitrary $N$. In this case there is equidistant position 
of the components
with different hypercharge, in view of Gell-Mann --- Okubo relations, and $Y^{max}=(N+3)/3$,
$Y^{min}=-(N+3)/6$, $\Delta Y= p+q =(N+3)/2$, and the mass splitting of this multiplet is
\be
\Delta^{tot}_{RR} M("\overline{10}") = m_K^2 \Gamma {N\Delta_Y \over (N+3)(N+9)} =
m_K^2 \Gamma {N\over 2(N+9)}.
\ee
Within BSM and its RO variant we have, without hyperfine splitting correction,
\be
\Delta^{tot}_{RO} M("\overline{10}") \simeq \Delta Y\omega_-\simeq {N(N+3)\over 16\Theta_K}(\mu -1)
\simeq {N+3\over 2N} m_K^2 \Gamma.
\ee
We cannot, however, to calculate the HFS correction in this case, because expression (\ref{hfsa}) is
not sufficient for this purpose.
To calculate the hyperfine correction for states with strangeness $S<1$ we should, in terms of
the quark model, make summation
of spins of nonstrange quarks, strange antiquark and several strange quarks, in correspondence with
strangeness $S$. This is more complicated problem to be solved starting from incident lagrangian.

To conclude this subsection, we obtained agreement between the RR and modified RO models in the 
total mass splitting of nonexotic baryon multiplets in two leading orders of $1/N$ expansion,
and for exotic multiplets - only in first leading order. The next order contributions in RO model
are not calculated yet. Anyway, since the expansion parameter is large, like $6/N$, the knowledge
of several terms of such expansion may be not so useful for extrapolation to the real $N=3$ world.
\section{Quark wave functions of pentaquarks}
The connection between chiral soliton models and the quark models of exotic states has been discussed
intensively, and different opinions have been revealed, from that both models are dual
\cite{manoh,vkuf},
or complementary to each other, to that they are essentially different, and predict different states:
in particular, in \cite{cld} the states were predicted which are absent in the simplest quantization scheme
of the chiral soliton models - the partners of states with different spin, but same flavor
quantum numbers, including isospin.
Here we show that some features of exotic baryons spectra obtained within
the chiral soliton approach can be illustrated in terms of the quark model, as it was shown at first
\cite{close} for the case of antidecuplet. Any model with $SU(3)$ flavor symmetry and its violation
in special way mimics the quark model in view of Gell-Mann --- Okubo type relations (section 2).
There are, however, some distinctions, mainly in the quantitative estimates of mass differences of
different diquarks and partners of exotic baryon states.

Under the {\it simple quark model} of baryons we mean the model where mass splittings within $SU(3)$
multiplets are defined mainly by difference between strange and nonstrange quark masses. It is
a common feature of phenomenological models discussed recently in connection with observation of
pentaquarks \cite{kl,jw}. Here we shall reserve a possibility that strange quark mass can be different
in different $SU(3)$ multiplets, as well as strange antiquark mass is different from the mass of
strange quark. There is nothing special in this assumption: even the effective masses of electrons
are slightly different in different atoms due to different binding energies. Strong interactions of
strange quarks and antiquarks with $(u,d)$ quarks are different, which can lead to considerable
difference of effective masses.

Under the {\it simplistic}, or {\it oversimplified quark model} we mean the model where strange quark and
antiquark masses are equal, as well as they are equal in different $SU(3)$ multiplets.
The striking property of exotic spectra within CSA is that the mass splitting within antidecuplet
in RR model, in the first order of perturbation theory for $N=3$ equals exactly to that of decuplet, 
as it follows from values of $C_S$ presented in Table 1,
therefore simplistic quark model contradicts the results of CSA for $N=3$.

As it follows from the formulas of the preceding section, the RO variant of the bound state model
in the leading in $1/N$ approximation corresponds to simple quark model, with the strange
quark mass
\be\label{qmass}
m_s \simeq m_K^2{\Gamma\over N},
\ee
which is of the order $N^0 \sim 1$ (as it follows from above results, the relation is rather
$m_s \simeq m_K^2\Gamma/( N+9)$, considerably smaller numerically for $N=3$). The antiflavor 
excitation energy $\omega_+$ is greater than $\omega_-$, so, one could decide that the effective 
mass of the strange antiquark is greater than the mass of the strange quark. Within the RR variant 
of the CSA the difference $\omega_+-\omega_-$ is reproduced by difference of rotational energies 
of different $SU(3)$ multiplets, due to difference of Casimir operators of exotic and nonexotic 
multiplets, and can be ascribed to the contribution
of the effective mass of additional quark-antiquark pair, $m_{q\bar q}\sim 1/\Theta_K$ 
(see, e.g. Appendix of \cite{ksh} and \cite{vk05}). Within the bound state model and its RO 
variant calculations of spectra of exotic multiplets (not only positive strangeness components) 
are absent still, as mentioned above.

Relation (\ref{qmass})
is in agreement with the known relation $m_s|<\bar q q>| \simeq F_K^2\mu_K^2/8$ \cite{indur}, with
the proper relation between quark condensate $<\bar q q>$ and $F_\pi^2/\Gamma$.
Sometime in the literature the relation is used to obtain $\Gamma$ or other quantities for
arbitrary $N$ from the value at $N=3$:  $\Gamma(N) = \Gamma(N=3)(N/3)$. We want to note here
that this is really arbitrary and not justified prescription, since any relation of the type
$\Gamma(N)=\Gamma(N=3)[(N+a)/(3+a)]$ with any real (positive) constant $a$ gives correct value
for $N=3$, but different at large $N$.
\subsection{Quark contents of exotic baryons in pentaquark approximation}
 We call $q$ the lightest quarks,
$u$, $d$, and $s$ denotes as usually the strange quark, ($c$, $b$ - the charmed or beauty quark).
We consider here the case of strangeness, the charmed or beautiful states can be obtained by
simple substitution $s \to c$, etc.

{\it Quark contents of antidecuplet.} First we recall that the minimal quark content of the 
components of $\{\bar{10}\}$-plet is, for $N=3$  \cite{close}:
\bea
&\Theta^+:& |\bar{10},2,0,0>\;\; =|uudd\bar{s}>; \nonumber\\
& N^*:& |\bar{10},1,1/2,-1/2>  =\; |udd (Q\bar{Q})_{N^{*0}}>,
|\bar{10},1,1/2,1/2> = |uud  (Q\bar{Q})_{N^{*+}}>; \nonumber \\
& \Sigma^*:&|\bar{10},0,1,-1>\; = |sdd (Q\bar{Q})_{\Sigma^{*-}}>,...\;,\;
|\bar{10},0,1,1> = |suu (Q\bar{Q})_{\Sigma^{*+}} >;\nonumber \\
& \Xi^*_{3/2}:&|\bar{10},-1,3/2,-3/2>\; =\; |ssd d\bar{u}>,\;...\;,\;
|\bar{10}, -1,3/2,3/2>= |ssu u\bar{d}>.
\eea
Here we use the notation $|N(p,q), Y, I, I_3>$ for the components of the multiplet 
$N(p,q)=(p+1)(q+1)(p+q+2)/2$ with
hypercharge $Y$, isospin $I$ and its third projection $I_3$.
The minimal quark content (i.e. the number of $u,\,d,\;s$ quarks or antiquarks) of manifestly exotic
states $\Theta^+$ and $\Xi^*_{3/2}$ is unique within pentaquark approximation, 
the condition for this is $I=(5+S)/2$ for $S\leq 0$, since the number of
nonstrange quarks and antiquarks equals to $5+S$ and each of them has isospin $1/2$.
This uniqueness of the quark contents allows to obtain the mass splitting within simple
quark model and to compare with results of the chiral soliton (rigid rotator version) model described
above.

In the  model  with $\bar{3}_F$ diquarks \cite{jw,close} the flavor part of the wave
function of $\Theta^+$ is made of two isoscalar diquarks:
\be
\label{Psi0}
\Psi_{\Theta^+} = {1\over 2}[u_1d_2-u_2d_1][u_3d_4-u_4d_3] \bar{s}
\ee
which corresponds exactly to isospin $I=0$. Other components of antidecuplet can be obtained
by action of $U$-spin, or $V$-spin and isospin operators ($U\,d =s,\;\; U\bar s = -\bar d$, etc., 
see e.g. \cite{close}).

The quark contents and the wave function of cryptoexotic states $N^*$ and $\Sigma^*$ depend on
the particular model:
 $(Q\bar{Q})_B = \alpha_B s\bar{s} +\beta_B u\bar{u} + \gamma_B d\bar{d}$ with coefficients $\alpha,\;
 \beta,\;\gamma$ depending not only on
particular baryon under consideration but also on the variant of the model and on mixing between
different $SU(3)$ multiplets. Within diquark model \cite{jw,close} one obtains
\be
\label{PsiN}
\Psi_{N^{*+}} = {1\over \sqrt 3}\left ([us]_{12}[ud]_{34}\bar s +[ud]_{12}[ud]_{34}\bar s-
 [ud]_{12}[ud]_{34}\bar d\right ),
\ee
with $[us]_{12}=(u_1s_2-u_2s_1)/\sqrt 2,$ and similarly for other cryptoexotic components 
of antidecuplet, see Table 3.

The wave function of the $\Xi$ -quartet does not contain $(s\bar{s})$
pair as a consequence of isotopic invariance: we can obtain components $\Xi^{*-}_{3/2},\;
\Xi^{*0}_{3/2};\;\Xi^{*+}_{3/2}$ from $\Xi^{*--}_{3/2}$ by acting operator $I^+$, and the $(s\bar{s})$
pair does not appear.

The upper component of antidecuplet $\Theta^+$ (see Fig. 1) contains one antiquark with the mass 
$m_{\bar s}$, the lower component, $\Xi_{3/2}$, 
contains two strange quarks with the mass $2m_s$, therefore, the whole splitting due to the
mass of the strange quark is $1\,m_s$, within simplistic model \cite{close}, and within
pentaquark approximation, of course. This should be compared with the total splitting
$3\,m_s$ for decuplet, where minimal content varies from $(qqq)$ for $\Delta$-isobar to
$(sss)$ for $\Omega$-hyperon.
The particular weight of $(s\bar{s})$ pair in intermediate components (with strangeness $0$ and $-1$)
depends on the assumption concerning  structure of their wave function. It can be different in
different models, e.g. diquark-diquark or diquark-triquark models and even for different
variants of the diquark model. In the model \cite{jw} the equidistant behaviour was obtained for
antidecuplet \cite{close}.
But such behaviour of antidecuplet spectrum does not follow in general from above
consideration \footnote{In the paper \cite{jw} the mixing between pentaquark octet and
antidecuplet was studied, but their mixing with the lowest baryon octet was neglected. Strong interactions
do not conserve the number of additional quark-antiquark pairs, therefore, this mixing takes place
inevitably and will push the states considered towards higher energies. The nonexotic octet and decuplet
of baryons should be included into consideration for selfconsistency of any model. The paper \cite{jenman}
contains a similar remark.}.

{\it Quark contents of $\{27\}$-plet. }
The $\{27\}$-plet has the upper $S=+1,\;I=1$ component with content $qqqq\bar{s}$ of mixed symmetry
and manifestly exotic components with $S=-1, \;I=2,\;\;S=-2,\;I=3/2$ and $S=-3,\;I=1$, the components
with $S=0,\;I=3/2$ or $I=1/2$ are cryptoexotic:
\bea
& \Theta_1:&  |27,2,1,-1> = |dddu\bar{s}>,\;...\;,\; |27,2,1,1> = |uuud\bar{s}>;\nonumber\\
&\Delta^*:& |27,1,3/2,-3/2>\; =\; |ddd (Q\bar{Q})_{\Delta^{*-}}>,\;...\;,\;
|27,1,3/2,3/2>\;=\;|uuu (Q\bar{Q})_{\Delta^{*++}}>;\nonumber\\
&\Sigma_2:&|27,0,2,-2>\;=\;|sddd\bar{u}>,\;...\;,\; |27,0,2,2>\;=\;|suuu\bar{d}>;\nonumber\\
&\Xi^{*}_{3/2}:& |27,-1,3/2,-3/2> = |ssdd\bar{u}>,\;...\;,\;|27,-1,3/2,3/2>=|ssuu\bar{d}>;\nonumber\\
& \Omega_1:& |27,-2,1,-1> = |sssd\bar{u}>,\;...\;,\;|27,-2,1,1> =|sssu\bar{d}>,\eea
so, the energy gap is $2m_s$ for $4$ units of strangeness, $m_s/2$ in average.
Evidently, the upper $S=+1,\; I=1$ component of $\{27\}$-plet, as well as $S=+1$ component of $\{35\}$-
plet cannot be obtained in the flavor antisymmetric diquark model \cite{jw}. The flavor symmetric
diquarks of the type $6_F$ (isovectors in the $S=0$ case) must be invoked for this purpose.

Indeed, if diquark is $\bar{3}_F$, then we have according to well known group-theoretical relation:
\be\label{333}
 \bar{3} \otimes \bar{3} \otimes \bar{3} = \overline{10} \oplus 8 \oplus 8 \oplus 1,
 \ee
and there appears only antidecuplet from known pentaquark states (Fig. 1), and two octets of baryons.
If one diquark is $\bar{3}$, and the other is $6_F$, we obtain
\be\label{633}
6 \otimes \bar{3} \otimes \bar{3} = (15 \oplus 3) \otimes \bar{3} =
27 \oplus 10 \oplus 8 \oplus 8 \oplus 1.
\ee
If both diquarks are $6_F$, then 
\be\label{663}
6 \otimes 6 \otimes \bar{3} = (15 \oplus 15 \oplus \bar{6}) \otimes \bar{3} =
35 \oplus 10 \oplus 27 \oplus 10 \oplus 8 \oplus \overline{10} \oplus 8.
\ee
So, in the latter case all known pentaquark states can be obtained
\footnote{For example, the $S=+1$ component of antidecuplet made of two isovector diquarks
is $\Psi_{\Theta^+}=[u_1u_2d_3d_4+ d_1d_2u_3u_4-{1\over 2}(u_1d_2+u_2d_1)(u_3d_4+d_3u_4)]\bar{s}$.
In the diquark-triquark model \cite{kl} the diquark within triquark is color-symmetric $(6_c)$ and
anti-triplet in flavor, so, this model should be modified to provide $\{27\}$- and $\{35\}$-plets
of pentaquarks}.

Let us denote $(q_1q_2)$ the flavor symmetric diquark, $6_F$ in flavor, with spin $J=1$ 
($\bar{3}_C$ in color).
Then realization of the wave function of $\{27\}$-plet of pentaquarks via diquarks is:
\be |27,2,1,1> = (u_1u_2)[u_3d_4]\bar{s}, \ee
other components can be obtained with the help of $U$-spin and isospin 
$I^{\pm}$ operators:
\bea &|27,1,3/2,3/2>& = (u_1u_2)\biggl[[u_3s_4]\bar{s}- [u_3d_4] \bar{d}\biggr]/\sqrt{2},\nonumber\\
&|27,0,2,2>& = -(u_1u_2)[u_3s_4]\bar{d}, \nonumber\\
& |27,-1,3/2,3/2>& = -(u_1s_2)[u_3s_4]\bar{d}, \nonumber\\
& |27,-2,1,1>& = (s_1s_2)[u_3s_4]\bar{d}, \eea
It follows that the weight of the $s\bar{s}$ pair within $S=0$ component is 1/2, therefore, the
contribution of the strange quark mass equals $m_s$ in this case, similar to $|27,2,1>$ state
The $S=-1,\;I=2$ components have content $sqq q\bar{q}$, from $sddd\bar{u}$ to $suuu\bar{d}$, and
it does not contain $s\bar{s}$-pair. Therefore, its mass contains $1\,m_s$, similar to $S=+1, \;I=1$
component see Table 3). Remarkably, that chiral soliton calculation provides very close results 
for masses of
$S=+1$ and $S=-1, \;I=2$ components of $\{27\}$-plet, (Table 2 and Fig. 4 of \cite{wk}): the
difference of masses equals $0.03\;GeV$, see Table 3 which is modification of Table 5
in \cite{vk05}.

The effect of configuration mixing is especially important for cryptoexotic components of
the antidecuplet ($Y=1$ and $0$) which mix with similar components of the lowest baryon octet.
As it is known from quantum mechanics, in this case mixing makes the splitting between
the octet and antidecuplet greater and pushes the upper state to higher energy.
The mixing of the manifestly exotic state $\Xi_{3/2} \in \{\overline{10}\}$ with corresponding 
component of $\{27\}-$plet pushes it down, as a result the total mass splitting within 
$\bar{10}$ becomes smaller due to mixing.

\begin{center}
\begin{tabular}{|l|l|l|l|l|l|}
\hline
$|\overline{10},2,0>$&$|\overline{10},1,1/2>$&$|\overline{10},0,1>$&$|\overline{10},-1,3/2>$& &\\
\hline
$m_{\bar s}$&$2m_{s\bar s}/3$ & $m_s+m_{s\bar s}/3$ & $2m_s$ & & \\
\hline
1503 & 1594 & 1684 & 1775 & &\\
1539 & 1661 & 1764 & 1786 & & \\
\hline
\hline
$|27,2,1>$&$|27,1,3/2>$&$|27,0,2>$&$|27,-1,3/2>$&$|27,-2,1>$&\\
\hline
$m_{\bar s}$&$m_{s\bar s}/2$ & $m_s$ & $2m_s$ & $3m_s$ & \\
\hline
1672 & 1692 & 1711 & 1828 & 1944 &\\
1688 & 1826 & 1718 & 1850 & 1987 &\\
\hline
\hline
$|35,2,2>$&$|35,1,5/2>$&$|35,0,2>$&$|35,-1,3/2>$&$|35,-2,1>$&$|35,-3,1/2>$ \\
\hline
$m_{\bar s}$&$0$ & $m_s$ & $2m_s$ & $3m_s$ & $4m_s$ \\
\hline
2091 & 1796 & 1910 & 2023 & 2136 & 2250 \\
2061 & 1792 & 1918 & 2046 & 2175 & 2306 \\
\hline
\end{tabular}
\end{center}
{\bf Table 3.} {\tenrm Masses of components of $\{\overline{10}\}$, and components with maximal 
isospin for $\{27\},\, J=3/2$ and $\{35\},\,J=5/2$-plets of exotic baryons (in $MeV$, the nucleon 
mass is input, $N=3$). The first line after notations of the
components shows the contribution of the strange quarks/antiquark masses within simple model, 
$m_{s\bar s}$ is the mass of the $s\bar s$ pair taken usually to the sum of masses of $s$ and $\bar s$
quarks.
The next line is the result of calculation without configuration mixing, the second line of numbers --- 
configuration
mixing included according to \cite{wk}. Calculations correspond to case $A$ of paper \cite{wk}:
$\Theta_K =2.84\,GeV^{-1},\; \Theta_\pi=5.61\,GeV^{-1},\;\Gamma =1.45\,GeV,$ which allowed to
obtain the mass of $\Theta^+$ hyperon close to the observed value $1.54\,GeV$.} \\

For the cryptoexotic component of $\{27\}-$plet the mixing effect is especially large: $\sim 20\%$
admixture of $\Delta$-isobar from decuplet pushes this component by additional $130\,MeV$ above
nucleon and makes it even higher in energy than nearest strange $\Sigma_2$ state.

{\it Quark contents of $35$-plet.}
The wave function of the $\{35\}$-plet, the largest multiplet of pentaquarks, is
symmetric in flavor indices of $4$ quarks.
The $I=2$ upper components of this multiplet has quark content $qqqq\bar{s}$, from $dddd\bar{s}$
to $uuuu\bar{s}$:
\be\Theta^{**}_2: |35, 2, 2, -2> = |dddd\bar{s}>; \;...\;,\; |35, 2, 2, 2> = |uuuu\bar{s}>
\ee
The intermediate components can be obtained easily by applying the isospin operators $I^+$ or $I^-$.
 Evidently, it has the largest possible isospin for the $S=+1$ pentaquarks.
The strange antiquark contribution into the mass equals  $m_{\bar s}$, obviously (and. 
$m_{\bar s}= m_s$ in simplistic model).
The $S=0$ components of $\{35\}$-plet with isospin $I=5/2$ has minimal content $qqqq\bar{q}$
(evidently, $I=5/2$ is the maximal possible value of isospin of any pentaquark):
$$\Delta_{5/2}: \;|35,1,5/2,-5/2>= |dddd\bar{u}>,\; ...\;,\; |35,1,5/2,5/2> = |uuuu\bar{d}>, $$
and do not contain strange quarks at all. By this reason, the $I=5/2,\; S=0$ component is the
lightest component of the $\{35\}$-plet, and has smaller strangeness content than nucleon and
$\Delta$, again in agreement with calculation within CSA \cite{wk}.
\begin{figure}
\label{quarkspl}
\begin{center}
\epsfig{figure=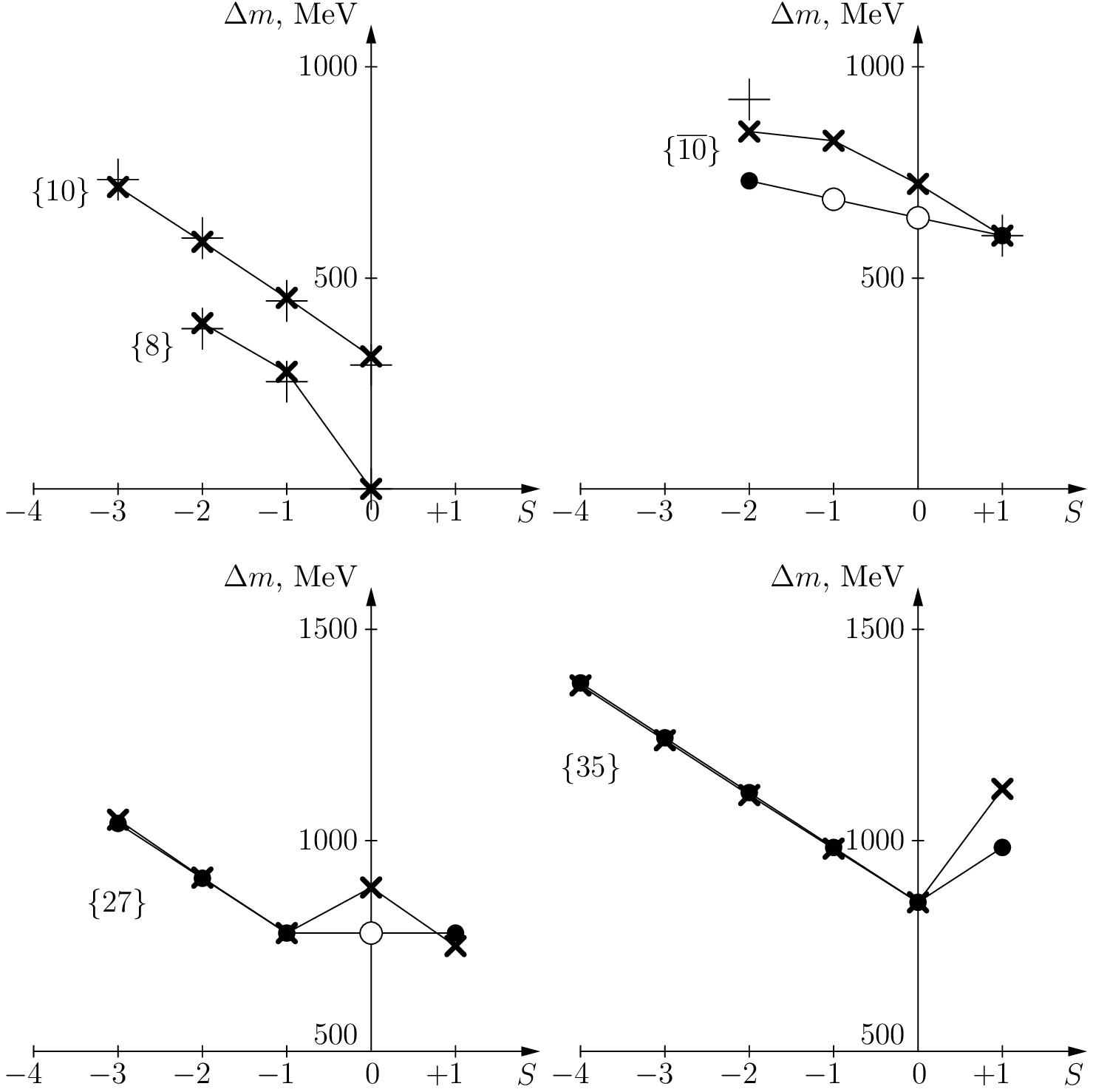,width=12cm,angle=0}
\protect\caption{\tenrm Schematic picture of the mass splittings within chiral soliton model $(N_c=3)$.
The upper left figure corresponds to the nonexotic octet and decuplet, the upper right one - to exotic
antidecuplet, the lower - to $\{27\}$-plet with spin $J=3/2$ and to $\{35\}$-plet $(J=5/2)$ of exotic 
baryons. Experimental data are
shown by direct crosses ${\bf +}$, position of states obtained within CSA with configuration mixing
is marked by ${\bf \times}$. The circles show position of states within the simplistic quark model
with $m_s=m_{\bar s} \simeq 130\,MeV$; full circles show manifestly exotic states with unique quark
contents and empty circles - cryptoexotic states. For the antidecuplet the fit is made for the state
with $S=1$, see also discussion in the text.}
\end{center}
\end{figure}
The components with $S=-1$, $S=-2$, etc. should contain strange quarks in the wave function:
\bea \Sigma_2: & |35,0,2,-2>& = |sddd\bar{u}>,\; ...\;,\; |35,0,2,2> = |suuu\bar{d}>;\nonumber\\
 \Xi^{**}_{3/2}:& |35, -1,3/2,-3/2>& = |ssdd\bar{u}>,\;...\;,\; |35,-1,3/2,3/2> = |ssuu\bar{d}>;
 \nonumber\\
\Omega_1^*:&|35, -2,1, -1>& = |sssd\bar{u}>,\;...\;,\; |35,-2,1,1 > = |sssu\bar{d}>;\nonumber\\
\Gamma {?}:&|35, -3,1/2,-1/2>& = |ssss\bar{u}>,\;\; |35,-3,1/2,1/2> = |ssss\bar{d}>, \eea
and there is no place for the $s\bar{s}$ pair \footnote{The notation $\Gamma$ for the $S=-4,\,I=1/2$
component of $\{35\}$-plet is not generally accepted, still.}. 
The 4-quark part of the wave function of the $\{35\}$-plet is symmetric in flavors and can be easily
made of two flavor symmetric $6_F$ diquarks, e.g. $\{ddds\}=(dd)(ds)+(ds)(dd),\; \{ddss\}=
(dd)(ss)+(ss)(dd)+(ds)(ds),$ etc.

The lowest $S=-4$, $I=1/2$ isodoublet has $4m_s$ contribution in the mass. As a result, we have
the mass gap $4m_s$
between $S=0,\;I=5/2$ state and $S=-4,\;I=1/2$ state: $1\,m_s$ for unit of strangeness. But the gap
between $S=+1$ and $S=-4$ components is only $3\,m_s$ for 5 units of strangeness, $3m_s/5$ for one
unit in average. The result of chiral soliton model calculation \cite{wk} is in rough agreement with
the mass splitting given by the above wave function with $m_s \simeq 130\; -\;140\;MeV$. All exotic
components of $\{35\}$-plet mix with components of higher irreducible representations 
$(\{64\}$-plet, etc) and slightly move down in energy
after mixing. Positions of states obtained within CSA are shown in Fig.2 with ${\bf \times}$.
Predictions of simplistic quark model with $m_s=m_{\bar s}=130\,MeV$ are shown with circles.
For $\{27\}$-plet the location of state with $S=-1$ is identified with that of CSA, same for $S=0$
component of the $\{35\}$-plet.

Summing up, within simplistic quark model we have the following hierarchy of the energy gaps per 
unit strangeness (in average) between highest and lowest components of the $SU(3)$ multiplets:
$m_s/3;\; m_s/2;\; 3m_s/5$ for $\{\overline{10}\}$, $\{27\}$ and $\{35\}$-plets, but the individual
splittings, in general, do not follow such simple law and are model dependent.
Obviously, this is in contradiction with CSA approach results, and we should allow the masses
of strange quarks be different within different $SU(3)$-multiplets.
Then the following relations take place, according to the results presented in Table 3 (configuration
mixing included):
\bea
\left[ 2m_s - m_{\bar s}\right]_{\overline{10}}&\simeq& 250\,MeV;\nonumber\\
\left[ m_s - m_{\bar s}\right]_{27} &\simeq& 30 \,MeV, \qquad [m_s]_{27}\simeq 135\,MeV;\nonumber\\
\left[ m_s\right]_{35} &\simeq& 130\,MeV;\qquad \left[m_{\bar s}\right]_{35}\simeq 270\,MeV.
\eea
Only one relation takes place for the antidecuplet, and if we assume that the mass of strange quark 
within the antidecuplet is close to that within higher multiplets, i.e. about $130-135\,MeV$, then 
strange antiquark within $\overline{10}$ should be very light, $\sim 10-20 \,MeV$ only. The strange 
antiquark is heavier within $\{27\}$-plet, about $100\, MeV$, and much heavier within $\{35\}$-plet.
Recall that now the observed mass splitting within antidecuplet is about $320\,MeV$, if the observed
$\Xi^{--}$ state \cite{alt} belongs to the antidecuplet, and not to the higher multiplet. To fit 
simplistic quark model, the splitting of the antidecuplet should be smaller, about $130-150\,MeV$, 
but this will be in disagreement with CSA. 

Some decrease of the "strange" (or kaonic) inertia $\Theta_K$ in comparison 
with the value used to obtain the numbers in Table 3 \cite{wk,vk05} would increase all masses of 
exotic states, but would not make much influence on the mass splittings inside of $SU(3)$ multiplets.
Experimental studies of exotic spectra could help in solving this problem, present situation with 
searches of baryons from higher $SU(3)$ multiplets has been discussed recently in \cite{azimov}.
\subsection{Diquarks mass difference estimate}
Comparison with results of chiral soliton approach allows to estimate the difference of the 
diquarks masses as well. 

In the rigid or soft rotator approximation there is contribution to the mass difference of the 
different $SU(3)$-multiplets due to different rotation energy (second order Casimir operators) of these
multiplets. For $\{27\}$- and $\{\overline{10}\}$-plets this difference is
\be
M^{rot}_{27,J=3/2} - M^{rot}_{\overline{10}} = {3\over 2\Theta_\pi} - {1\over 2\Theta_K}.
\ee
This difference can be naturally ascribed to the difference of effective masses of $6_F$- and 
$\bar 3_F$ diquarks (see (\ref{333}) and (\ref{633}) above). This quantity is about $100\,MeV$,
more precisely, $91\,MeV$ if we take the same values of moments of inertia, as in Table 3.
The difference of rotational energies of $\{35\}$-plet which contains two $6_F$ diquarks (see
(\ref{663})) and $\{27\}$-plet is
\be
M^{rot}_{35,J=5/2} - M^{rot}_{27,J=3/2} = {5\over 2\Theta_\pi} - {1\over 2\Theta_K}.
\ee
Numerically this is considerably greater than in the former case, about $270\,MeV$. 
The real picture may be considerably more complicated: besides effective masses of diquarks
the interaction energy between different diquarks can be substantially different. 
This means that there is no simple additivity of the diquark masses within topological soliton approach.
Roughly, we can conclude however that the mass difference between $6_F$ and $\bar 3_F$ diquarks 
is between $100$ and $270\,MeV$, the latter value is close to the estimate given, e.g. in \cite{fw}.

Consideration of charmed or beautiful states can be made in close analogy with that for strangeness.
One could consider $SU(4)\; (u,d,c,s)$ or even $SU(5)\; (u,d,c,s,b)$ symmetry, but since this
symmetry is
badly violated, it has not much significance for practical use. Instead, 
the $(u,d,c)$ and $(u,d,b)$ $SU(3)$ symmetry groups are often considered.
The $\{35\}$-plet is again remarkable: within $SU(4)$ it should belong to the most symmetric 
$\{120\}$-plet
which can be described by spinor $T^{iklm}_r$, $(i,k...r=u,d,s,c)$, corresponding Young tableau is
$(4,0,1)$; within $SU(5)\;(u,d,s,c,b)$ it belongs to $\{315\}$-plet with Young tableau $(4,0,0,1)$.
The $S=0$, or $c=0$, or $b=0$ components of $\{35\}$-plets which do not contain $s\bar{s}$ or 
$c\bar{c}$, or $b\bar{b}$ in the wave function is a common component of the $\{35\}$-plets in 
each of $SU(3)$ groups, which is remarkable property of this $I=5/2,\;S=c=b=0$ state consisting of
light $u,\,d$ quarks only.
\section{Partners of exotic states with different values of spin.}
Within quark models there are partners of states with same flavor quantum numbers (strangeness and 
isospin), but with different value of spin \cite{cld}. Existence of partners of exotic baryons
has been demonstrated and discussed also in \cite{cole} in large $N_c$ QCD. At the same time, within 
CSA the value of spin equals to the value of "right" isospin. as a result of the "hedgehog" nature 
of the basic classical configuration.
\begin{figure}
\label{anti35}
\begin{center}
\epsfig{figure=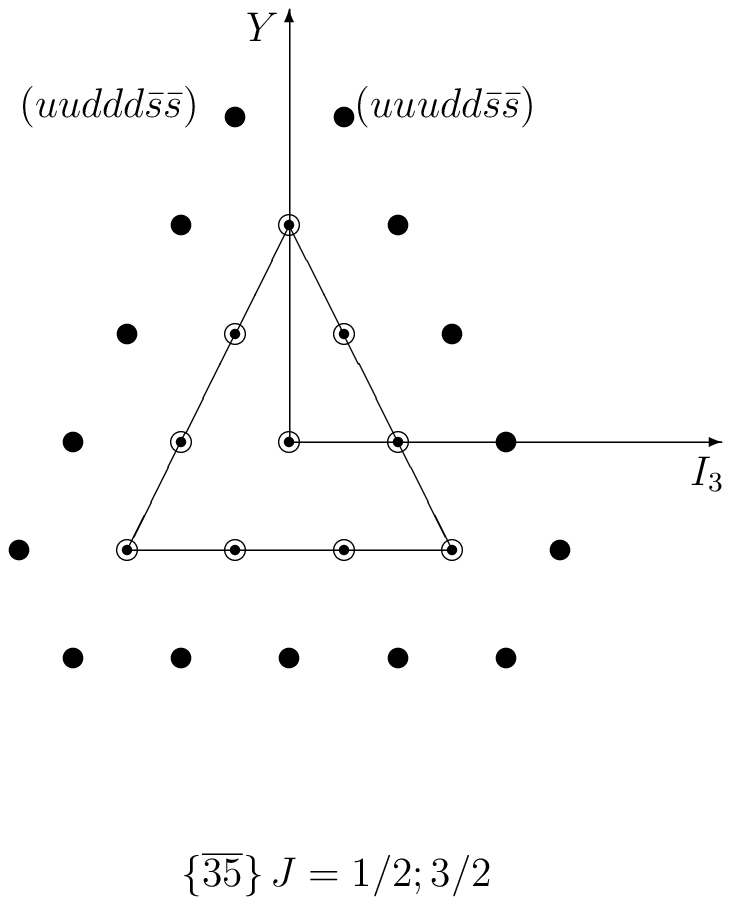,width=9cm,angle=0}
\protect\caption{\tenrm Partners of the components of the exotic antidecuplet located within
 $\{\overline{35}\}$-plet.}
\end{center} 
\end{figure}
A natural question is: where are such partners within CSA, if they exist at all?
The answer is that they are present as well, although belong to different $SU(3)$-multiplets.
Here we give one simple
example: the $J^P=3/2^+$ partner of antidecuplet with spin $J=1/2$ found its
 place within $\{\overline{35}\}$-plet $(p,q)=(1,4)$ (septuquark or heptaquark), as shown in
Fig. 3. The mass  of this state is considerably greater due to a large difference of the 
Casimir operators $C_2(SU_3)$:
 \be \label{diffpart}
 \Delta M^{rot}_{\bar{35}-\bar{10}}
  = M(\overline{35},J=3/2) - M(\overline{10})={3\over 2\Theta_K}+{3\over 2\Theta_\pi}
 \ee
which is about $750-800\,MeV$, greater than several tens of $MeV$ obtained in \cite{cld}.
The spectrum of these states for some reasonable values of model parameters is given in Table 6 of
\cite{vk05}, and we shall not reproduce it here.

There are also partners of nonexotic baryon states. For example,
the $J^P=5/2^+$ partners of the decuplet $(J^P=3/2^+)$ are
contained within $\{35\}$-plet $(4,1)$, the difference of rotational energies is
\be \label{diffpart2}
 \Delta M^{rot}_{35-10}
  = M(35,J=5/2) - M(10,J=3/2)={1\over 2\Theta_K}+{5\over 2\Theta_\pi}
 \ee
which is about $600 - 700\,MeV$.
Analog with $J^P=3/2^+$ of the baryon octet ($J^P=1/2^+$)
 is contained within $\{27\}$-plet and has energy by $0.7 - 0.85\,GeV$ greater than masses of the 
 lowest octet. Moreover, for any baryon multiplet one can find partners with greater value of 
 spin within some $SU(3)$ multiplet with other (greater, as a rule) values of $(p,q)$. 
So, all partners noted are present in the CSA as well, but have considerably greater energy.
It was assumed in \cite{jw} that the $J=3/2$ partners of exotic baryon states have considerably
greater energy than the $J=1/2$ ground states, and estimates made here can be considered as
justification of this assumption within chiral soliton model.

Another kind of partners are states with same value of spin (and parity), but another
value of isospin. Such partners are absent within multiplets of nonexotic baryons 
(octet and decuplet) and for the antidecuplet, but exist for complicated multiplets, 
$\{27\}$- and
$\{35\}$-plets. The mass difference between such partners is due to FSB contributions in
(\ref{mf}), since rotational energy is the same, and is usually within few tens of $MeV$.
\section{Conclusions}
Calculations of the strangeness contents of exotic baryons, performed in present paper at arbitrary 
$N_c$ for the first time, have shown that the expansion parameter for this quantity
is large and increases for exotic states in comparison with nonexotic \cite{vk05,kleko}. 
There is common agreement that the rigid rotator model and the bound state approach provide
the same results in the limit $N_c\to \infty$, but there is crucial
difference in the following in $1/N_c$-expansion terms for different variants of the model --- 
rigid rotator variant and bound state model. There is a way to reach coincidence in the
next to leading in $1/N_c$ expansion terms by means of appropriate resolution of some ambiguities
in the BSM \cite{kleko}, but it is valid for large enough $N_c$, only. This makes questionable the 
possibility of extrapolation from the large $N_c$ to real $N_c=3$ world, and provides grounds for 
scepticism that conclusions made in the limit $N_c\to \infty$ - e.g. concerning existence or 
nonexistence of exotic baryon resonances - are valid in realistic case $N_c=3$ \cite{vk05}. This 
problem has been noted recently also for the quantities different from spectra of baryons, e.g. 
for widths of exotic resonances \cite{CHL,hw2}.
The existence of pentaquark states by itself seems to be without any doubt within CSA \cite{vk05,hw2}, 
although prediction of their particular properties like mass and width contains considerable 
uncertainties, 
and some kind of phenomenological extrapolation should be and has been made for this purpose, 
as e.g. in \cite{wk,ellis,trampetic}.

We have considered also some general properties of the pentaquark wave functions, mainly their 
quark contents for the realistic $N_c=3$ case.
The peculiarity of manifestly exotic states is that their quark contents are model independent
(within the pentaquark approximation), whereas the contents as well as wave functions of
cryptoexotic states depend on the particular variant of the model. 

The mass splittings within multiplets of pentaquarks (negative strangeness) expected within simple
quark model are reproduced in chiral soliton model (its rigid rotator variant), due to Gell-Mann ---
Okubo relations. In particular, the
lightest component of $\{35\}$-plet, the $\Delta_{5/2}$, which does not contain strange quarks or 
antiquarks within pentaquark approximation, is the lightest one in chiral soliton model as well.
For positive strangeness components of pentaquarks multiplets the link between CSM and QM requires
strong dependence of effective strange antiquark mass on the $SU(3)$-multiplet to which pentaquark
belongs. Configuration mixing pushes the spectra towards simplistic model --- nice property which
reasons are not clear yet.

The bound state model (its RO variant), in the leading in $1/N_c$ order, corresponds
to {\it simplistic} variant of the quark model with the unique value of the strange quark (antiquark)
mass, $m_s \simeq m_K^2\Gamma/N$. The next to leading order corrections for spectrum of 
exotic baryons with $(S<1)$ and correspondence with the simple quark model still remain to 
be investigated.

The partners of baryons multiplets with different $J$, discussed in the literature \cite{cld,cole},
for example the $J^P=3/2^+$ partner of the $1/2^+$ antidecuplet \cite{cld}, exist
within chiral soliton models as well \cite{vk05}. They are the parts of higher multiplets and have 
considerably greater energy than the states with the lowest value of spin.

In view of considerable theoretical uncertainties connected, in particular, with the problem of
extrapolation to realistic value of $N_c$, experimental searches for pentaquark states could be
decisive. Even if the existence of narrow pentaquarks is not confirmed, they can exist as broader
resonances of higher mass, and their studies will be useful for checking and development of 
theoretical ideas \footnote{Situation with observation of the $\Theta^+$ pentaquark state does not
become less dramatic: recently CLAS collaboration disavowed their previous result on $\Theta^+$
photoproduction on deuterons \cite{clas}, whereas DIANA collaboration reinforced their
result on $\Theta^+$ production by kaons in Xe chamber \cite{diana}.}.
\section{Acknowledgements}
The $SU(3)$ configuration mixing codes arranged by Bernd Schwesinger and Hans Walliser have been
used for checking analytical results at $N_c=3$.
VBK is indebted to Igor Klebanov for permission to use unpublished notes \cite{kleko}, to G.Holzwarth, 
J.Trampetic, H.Walliser and H.Weigel for E-mail conversations and discussions, and to Ya.Azimov, 
T.Cohen, K.Hicks, M.Karliner, H.Lipkin, M.Praszalowicz and other participants of Pentaquark05 
Workshop for useful discussions.
Results of this paper have been presented in parts at Pentaquark05 Workshop, Jefferson Lab.,
20-22 October 2005 and at PANIC-05, Santa Fe, New Mexico, 24-28 October 2005.
The work supported in part by RFBR, Grant No. 05-02-27072-z.

\section{Appendix: Wave functions of baryons in the $SU(3)$ configuration space for arbitrary number
of colors.}
In the rigid rotator quantization scheme the wave functions of baryon states are some combinations
of the $SU(3)$ Wigner D-functions. Such functions are quite well known for the case of $N_c=3$ and
for the octet and decuplet of baryons \cite{deSwart,Weigel}.
Here we present these functions for arbitrary number of colors and for exotic baryon multiplets,
since they are still absent in the literature.
As in \cite{deSwart,Weigel}, we have:
\be \label{psi}
\Psi(Y,I,I_3;Y_R,J,J_3)=\sum_{M_L} D^{I*}_{I_3,M_L}(\alpha,\beta,\gamma) f^{Y,I;Y_R,J}_{M_L,M_R}(\nu)
D^{J*}_{M_R,-J_3}(\alpha',\beta',\gamma') exp(iY_R\rho),
\ee
where $D^I_{M_1,M_2}$ are the well known $SU(2)$ Wigner functions, right hypercharge $Y_R=N/3$ and
$Y'_R=1$ for the case of baryons
we consider here, right isospin $I_R=J$, spin of the baryon state, due to the hedgehog structure of
the classical $B=1$ configuration, $M_R=M_L+(Y_R-Y)/2$ due to the properties of 
the $\lambda_4$ rotations. There are obvious restrictions $-I \leq M_L \leq I$, and $-J\leq M_R \leq J$,
and this leaves in the sum (\ref{psi}) few allowed terms. When the isospin of the state equals  $I=0$,
only one term is present in (\ref{psi}).
Nontrivial $\nu$ dependence is contained in the function $f^{Y,I;Y_R,J}_{M_L,M_R}(\nu)$ only, which
we present here. For the sake of brevity we label it further as $f_{M_L}$, since other labels can be
obtained easily, and we use notation $Q_{ikl...}=\sqrt{(N+i)(N+k)(N+l)...}$ for arbitrary
integers $i,k,l...$, some of them can be negative.\\

{\it "Antidecuplet": $[p,q]=[0,(N_c+3)/2]$}
\be
\Theta^+: \;\;
f_0=f_{0,-1/2}^{2,0;1,1/2}= {Q_{3,5,7} \over 4} s_\nu c_\nu^{(N+1)/2},
\ee
$$ Q_{3,5,7} = \sqrt{(N+3)(N+5)(N+7)}; $$
\be
N^*: \;\;
f_{-1/2} = f_{-1/2,-1/2}^{1,1/2;1,1/2}=\frac{Q_{5,7}}{8} (2-(N+3)s_\nu^2) c_\nu^{(N-1)/2}, \;
f_{1/2} = \frac{Q_{5,7}}{4} c_\nu^{(N+1)/2},
\ee
\be
\Sigma^{**}: \;\;
f_{-1} = \frac{Q_{1,5,7}}{8\sqrt6} s_\nu (4-(N+3)s_\nu^2) c_\nu^{(N-3)/2}, \;
f_0 = \frac{Q_{1,5,7}}{4\sqrt3} s_\nu c_\nu^{(N-1)/2},
\ee
\be
\Xi_{3/2}: \;\;
f_{-3/2} = \frac{Q_{-1,1,5,7}}{32\sqrt3} s_\nu^2 (6-(N+3)s_\nu^2) c_\nu^{(N-5)/2}, \;
f_{-1/2} = \frac{Q_{-1,1,5,7}}{16} s_\nu^2 c_\nu^{(N-3)/2}.
\ee

For each baryon state functions $f(\nu)$ are normalized according to \cite{deSwart}
\be
\int \left(\sum_m f_m^2\right) s_\nu^2 d s_\nu^2 =1.
\ee
The orthogonality conditions of wave functions of states with the same spin, strangeness and 
isospin, but from different $SU(3)$-multiplets, take the form, in view of orthogonality of
different $SU(2)$ $D$-functions:
\be
\int \left(\sum_m f_m g_m\right) s_\nu^2 d s_\nu^2 = 0,
\ee
which can be easily verified using wave functions given here.\\

{\it "27-plet": $[p,q]=[2,(N_c+1)/2]$}
\be
\Theta_1: \;\;
f_{-1} = \frac{Q_{1,3,9}}{4\sqrt2} s_\nu c_\nu^{(N-1)/2}, \;
f_0 = \frac{Q_{1,3,9}}{4\sqrt3} s_\nu c_\nu^{(N+1)/2}, \;
f_1 = \frac{Q_{1,3,9}}{4\sqrt6} s_\nu c_\nu^{(N+3)/2},
\ee
$$ \Delta^*: \;\;
f_{-3/2} = \frac{\sqrt3Q_{3,9}}{48} (6-3(N+1)s_\nu^2) c_\nu^{(N-3)/2}, \;
f_{-1/2} = \frac{\sqrt3Q_{3,9}}{48} (6-2(N+1)s_\nu^2) c_\nu^{(N-1)/2}, $$
\be
f_{1/2} = \frac{\sqrt3Q_{3,9}}{48} (6-(N+1)s_\nu^2) c_\nu^{(N+1)/2}, \;
f_{3/2} = \frac{\sqrt3Q_{3,9}}{8} c_\nu^{(N+3)/2},
\ee
$$ \Sigma_2: \;\;
f_{-2} = \frac{Q_{-1,3,9}}{16\sqrt{15}} s_\nu (12-3(N+1)s_\nu^2) c_\nu^{(N-5)/2}, \;
f_{-1} = \frac{Q_{-1,3,9}}{32\sqrt5} s_\nu (12-2(N+1)s_\nu^2) c_\nu^{(N-3)/2}, $$
\be
f_0 = \frac{Q_{-1,3,9}}{16\sqrt{30}} s_\nu (12-(N+1)s_\nu^2) c_\nu^{(N-1)/2}, \;
f_1 = \frac{3Q_{-1,3,9}}{8\sqrt{15}} s_\nu c_\nu^{(N+1)/2},
\ee
$$ \Xi_{3/2}^*: \;\;
f_{-3/2} = \frac{Q_{-1,3,7,9}}{64\sqrt5} s_\nu^2 (8-2(N+1)s_\nu^2) c_\nu^{(N-5)/2}, \;
f_{-1/2} = \frac{Q_{-1,3,7,9}}{32\sqrt{15}} s_\nu^2 (8-(N+1)s_\nu^2) c_\nu^{(N-3)/2}, $$
\be
f_{1/2} = \frac{Q_{-1,3,7,9}}{8\sqrt5} s_\nu^2 c_\nu^{(N-1)/2},
\ee
\be
\Omega_1: \;\;
f_{-1} = \frac{Q_{-1,3,5,7,9}}{64\sqrt{15}} s_\nu^3 (4-(N+1)s_\nu^2) c_\nu^{(N-5)/2}, \;
f_0 = \frac{Q_{-1,3,5,7,9}}{16\sqrt{10}} s_\nu^3 c_\nu^{(N-3)/2},
\ee
\\

{\it "35-plet": $[p,q]=[4,(N_c-1)/2]$}
$$ \Theta_2: \;\;
f_{-2} = \frac{Q_{-1,1,11}}{4\sqrt3} s_\nu c_\nu^{(N-3)/2}, \;
f_{-1} = \frac{Q_{-1,1,11}}{2\sqrt{15}} s_\nu c_\nu^{(N-1)/2}, \;
f_0 = \frac{Q_{-1,1,11}}{4\sqrt5} s_\nu c_\nu^{(N+1)/2}, $$
\be
f_1 = \frac{Q_{-1,1,11}}{2\sqrt{30}} s_\nu c_\nu^{(N+3)/2}, \;
f_2 = \frac{Q_{-1,1,11}}{4\sqrt{15}} s_\nu c_\nu^{(N+5)/2},
\ee
$$ \Delta_{5/2}: \;\;
f_{-5/2} = \frac{\sqrt5Q_{1,11}}{120} (10-5(N-1)s_\nu^2) c_\nu^{(N-5)/2}, \;
f_{-3/2} = \frac{\sqrt5Q_{1,11}}{120} (10-4(N-1)s_\nu^2) c_\nu^{(N-3)/2}, $$
\be
f_{-1/2} = \frac{\sqrt5Q_{1,11}}{120} (10-3(N-1)s_\nu^2) c_\nu^{(N-1)/2}, \;
f_{1/2}  = \frac{\sqrt5Q_{1,11}}{120} (10-2(N-1)s_\nu^2) c_\nu^{(N+1)/2},
\ee
$$ f_{3/2}  = \frac{\sqrt5Q_{1,11}}{120} (10- (N-1)s_\nu^2) c_\nu^{(N+3)/2}, \;
f_{5/2}  = \frac{\sqrt5Q_{1,11}}{12} c_\nu^{(N+5)/2}, $$
$$ \Sigma_2^*: \;\;
f_{-2} = \frac{Q_{1,9,11}}{48\sqrt{10}}  s_\nu (8-4(N-1)s_\nu^2) c_\nu^{(N-5)/2}, \;
f_{-1} = \frac{Q_{1,9,11}}{48\sqrt5} s_\nu (8-3(N-1)s_\nu^2) c_\nu^{(N-3)/2}, $$
\be
f_0 = \frac{Q_{1,9,11}}{16\sqrt{30}} s_\nu (8-2(N-1)s_\nu^2) c_\nu^{(N-1)/2}, \;
f_1 = \frac{Q_{1,9,11}}{24\sqrt{10}} s_\nu (8-(N-1)s_\nu^2) c_\nu^{(N+1)/2},
\ee
$$ f_2 = \frac{Q_{1,9,11}}{6\sqrt2} s_\nu c_\nu^{(N+3)/2}, $$
$$ \Xi_{3/2}^{**}: \;\;
f_{-3/2} = \frac{Q_{1,7,9,11}}{96\sqrt{15}} s_\nu^2 (6-3(N-1)s_\nu^2) c_\nu^{(N-5)/2}, \;
f_{-1/2} = \frac{Q_{1,7,9,11}}{96\sqrt5} s_\nu^2 (6-2(N-1)s_\nu^2) c_\nu^{(N-3)/2}, $$
\be
f_{1/2} = \frac{Q_{1,7,9,11}}{48\sqrt{10}} s_\nu^2 (6-(N-1)s_\nu^2) c_\nu^{(N-1)/2}, \;
f_{3/2} = \frac{Q_{1,7,9,11}}{8\sqrt6} s_\nu^2 c_\nu^{(N+1)/2},
\ee
$$ \Omega_1^*: \;\;
f_{-1} = \frac{Q_{1,5,7,9,11}}{192\sqrt{15}} s_\nu^3 (4-2(N-1)s_\nu^2) c_\nu^{(N-5)/2}, \;
f_0 = \frac{Q_{1,5,7,9,11}}{96\sqrt{15}} s_\nu^3 (4-(N-1)s_\nu^2) c_\nu^{(N-3)/2}, $$
\be
f_1 = \frac{Q_{1,5,7,9,11}}{24\sqrt6} s_\nu^3 c_\nu^{(N-1)/2},
\ee
\be
\Gamma_{1/2}: \;\;
f_{-1/2} = \frac{Q_{1,3,5,7,9,11}}{192\sqrt{30}} s_\nu^4 (2-(N-1)s_\nu^2) c_\nu^{(N-5)/2}, \;
f_{1/2} = \frac{Q_{1,3,5,7,9,11}}{96\sqrt6} s_\nu^4 c_\nu^{(N-3)/2}.
\ee

Wave functions of other states presented in Tables 1,2, and also states with another possible value
of spin have been obtained as well, but we shall not give them here for the sake of brevity.
\bigskip
\tenrm

\end{document}